\documentclass[aps,pra,twocolumn,superscriptaddress,longbibliography,footinbib]{revtex4-1}  %

\usepackage{graphicx} 
\usepackage{float} 
\usepackage{dcolumn} 
\usepackage{bm,color}
\usepackage{amsmath,amssymb,dsfont,amstext,amsfonts}
\usepackage{extarrows}
\usepackage{xcolor}
\usepackage{wasysym}
\usepackage{mathtools}
\usepackage{bbold}
\usepackage[export]{adjustbox}
\usepackage{mathdots}
\usepackage{siunitx}
\usepackage[colorlinks=true,linkcolor=blue,urlcolor=blue,citecolor=blue]{hyperref}
\usepackage{mathrsfs}

\usepackage[normalem]{ulem} 
\usepackage{color}

\usepackage{multirow}
\usepackage[T1]{fontenc}
\usepackage{lmodern}

\usepackage{xspace}

\newcommand*{\ie}{\textit{i.e.}\@\xspace}

\begin{document}
\title{Temperature effects in topological insulators of transition metal dichalcogenide monolayers}
\author{Siyu Chen}
\email{sc2090@cam.ac.uk}
\affiliation{TCM Group, Cavendish Laboratory, University of Cambridge,
J. J. Thomson Avenue, Cambridge CB3 0HE, United Kingdom}
\author{Isaac J. Parker}
\affiliation{Gonville and Caius College, 
University of Cambridge, Trinity Street, Cambridge CB2 1TA, United Kingdom}
\author{Bartomeu Monserrat}
\email{bm418@cam.ac.uk}
\affiliation{TCM Group, Cavendish Laboratory, University of Cambridge,
J. J. Thomson Avenue, Cambridge CB3 0HE, United Kingdom}
\affiliation{Department of Materials Science and Metallurgy, University of Cambridge, 27 Charles Babbage Road, Cambridge CB3 0FS, United Kingdom}

\begin{abstract}
We investigate the role of temperature on the topological insulating state of metal dichalcogenide monolayers, 1T$^\prime$-MX$_2$ (M=W, Mo and X=S, Se). Using first principles calculations based on density functional theory, we consider three temperature-related contributions to the topological band gap: electrons coupling with short-wavelength phonons, with long-wavelength phonons \textit{via} Fröhlich coupling, and thermal expansion. We find that electron-phonon coupling promotes the topology of the electronic structures of all 1T$^\prime$-MX$_2$ monolayers, while thermal expansion acts as a counteracting effect. Additionally, we derive the band renormalization from Fröhlich coupling in the two-dimensional context and observe its relatively modest contribution to 1T$^\prime$-MX$_2$ monolayers. Finally, we present a simplified physical picture to understand the ``inverse Varshni'' effect driven by band inversion in topological insulators. Our work reveals that, among the four 1T$^\prime$-MX$_2$ studied monolayers, MoSe$_2$ is a promising candidate for room temperature applications as its band gap displays remarkable resilience against thermal expansion, while the topological order of WS$_2$ can be tuned under the combined influence of strain and temperature. Both materials represent novel examples of temperature promoted topological insulators.
\end{abstract}
\maketitle

\section{Introduction}
In condensed matter physics, the unique electronic properties of topological insulators have sparked immense interest, captivating the scientific community with their potential to redefine the landscape of electronic and quantum computing technologies. Over the past two decades, significant advancements have been made in unravelling the fundamental features of topological insulators and in the manipulation of their distinctive properties~\cite{kane2005_1, kane2005_2, bernevig2006, hasan2010, qi2011, bansil2016, armitage2018, lv2021}.

The manipulation of topologically non-trivial band gaps has emerged as a highly active research direction. These manipulations have been achieved through various means, including alterations in chemical composition~\citep{hsieh2008, xu2011}, the application of external pressure~\citep{xi2013, bera2013}, the imposition of mechanical strain~\citep{liu2014, liu2016, lin2017_a}, and the influence of electromagnetic fields~\citep{kim2012, zhang2013, lin2017_b}. Recent attention within the scientific community has shifted towards exploring the intriguing possibility of controlling the topological characteristics of these materials through temperature, both in theory~\cite{garate2013, saha2014, monserrat2016, antonius2016, monserrat2019, peng2019, brousseau2020, querales2020, chen2022, marrazzo2023} and experiments~\cite{wiedmann2015, wojek2015, thirupathaiah2017, berger2018, chen2019top, regmi2020, estyunin2020, jiang2023, mohelsky2023}. 

Despite this burgeoning interest, a comprehensive examination of the interplay between temperature and the unique electronic states of topological insulators remains a relatively under-explored frontier. So far, most works indicate that increasing temperature suppresses the topological phase, in the sense that the topological band gap decreases with increasing temperature~\cite{monserrat2016, monserrat2019, brousseau2020, querales2020, jiang2023, mohelsky2023, marrazzo2023}. In particular, thermal expansion is often responsible for the ultimate suppression of topological order, as it drives the system towards the atomic limit. This suggests that a generic topological phase diagram as a function of temperature is one with a low temperature topological phase and a high temperature normal phase. Indeed, several reports describe the reduction of the topological band gap with increasing temperature culminating in a transition from a topological (crystalline) insulator into a normal insulator~\cite{wojek2015,monserrat2016, brousseau2020, querales2020} or from a strong topological insulator into a weak topological insulator~\cite{monserrat2019, jiang2023, mohelsky2023}. It has been argued that electron-phonon interactions can promote a topological insulating phase with increasing temperature in BiTlS~\cite{antonius2016} and bismuthene~\cite{chen2022}, but these works neglect thermal expansion so the overall temperature dependence in these materials remains an open question. As far as we are aware, the only compound in which temperature has been shown to promote topological order, specifically a topological semimetallic phase, is PbO$_2$~\cite{peng2019}, and there are no examples in which increasing temperature promotes a topological insulating phase.

Motivated by the paucity of reports on materials where temperature promotes topological insulating states, in this work we systematically study the interplay between temperature and topology in four transition metal dichalcogenide (TMD) monolayers 1T$^\prime$-MX$_2$ (where M=W, Mo and X=Se and S). The motivation behind the choice of monolayer TMDs for our study is two-fold. First, monolayer materials often exhibit unconventional thermal effects, including weak or negative thermal expansion, which may minimize the detrimental effects of thermal expansion on topological order observed in most compounds. Second, these specific TMD monolayers have been shown to be a realization of topological insulators at 0\,K~\cite{qian2014, tang2017quantum}, but their behavior at finite temperature is still an open question. 

Our investigation illustrates the pivotal role played by electron-phonon coupling and thermal expansion in influencing the topology of the electronic structures of 1T$^\prime$-MX$_2$ monolayers. Particularly noteworthy are MoSe$_2$ and WS$_2$, which stand out as the first example of temperature promoted topological insulators. The former emerges as a promising candidate for room temperature applications, owing to its remarkable resistance to thermal expansion, while the latter exhibits a tunable topological behavior when subjected to the combined influence of strain and temperature. Overall, our findings enrich the family of materials where temperature can promote topology.

The paper is organized as follows. In Section \ref{sec: ii}, we present the theory of finite temperature band structures and its first principles implementation in the context of the finite difference method. In Section \ref{sec: iii}, we apply the method to investigate 1T$^\prime$-MX$_2$, showcasing the competition between electron-phonon coupling and thermal expansion. In Section \ref{sec: iv}, we propose a simple model that aids in understanding the circumstances under which electron-phonon coupling can promote the topological phase in the presence of significant band inversion. In Section \ref{sec: v}, we show the topological order of WS$_2$ can be manipulated through a combination of strain and temperature, illustrating a transition from a low temperature normal to a high temperature topological phase. Finally, we summarize the contributions and findings of our research in Section \ref{sec: vi}.

\section{temperature dependent topological band gap}
\label{sec: ii}
A topological phase transition induced by temperature from a topological insulator to a trivial insulator (or the other way around) requires the bulk band gap to re-order. In the case of centrosymmetric materials, the band re-ordering is mediated by a gapless Dirac semimetal phase~\cite{murakami2007}, where the Dirac cone is located at one of the time reversal invariant momenta points. Therefore, the key to exploring the role of temperature in topological insulators is to consider the temperature dependence of certain energy eigenvalues.

The temperature dependent eigenenergy, $E_{n\boldsymbol{k}}(T)$,  of a single electron with band index $n$ and wavevector $\boldsymbol{k}$ can be approximated by the sum of two independent contributions: the renormalization of the eigenenergy by electron-phonon interactions, $\Delta E^{\mathrm{EP}}_{n\boldsymbol{k}}(T)$; and by thermal expansion, $\Delta E^{\mathrm{TE}}_{n\boldsymbol{k}}(T)$. For practical calculation reasons, the former needs to be further decomposed into the short- and long-wavelength phonon contributions, $\Delta E^{\mathrm{S}}_{n\boldsymbol{k}}(T)$, and $\Delta E^{\mathrm{L}}_{n\boldsymbol{k}}(T)$, each treated separately. Overall, the total shift of the electron eigenenergy at temperature $T$ reads
\begin{equation}
\Delta E_{n\boldsymbol{k}}(T) = \Delta E^{\mathrm{S}}_{n\boldsymbol{k}}(T) + \Delta E^{\mathrm{L}}_{n\boldsymbol{k}}(T) + \Delta E^{\mathrm{TE}}_{n\boldsymbol{k}}(T).
\end{equation}
In the following subsections, we discuss each contribution in turn. We use atomic units, $\hslash=e=m_{\mathrm{el}}=4 \pi \epsilon_{0}=1 $, unless otherwise stated.

\subsection{Band gap renormalized by short-wavelength phonons}
The investigation of band renormalization through coupling electrons to short-wavelength phonons has a long history, culminating in the well-established Allen-Heine-Cardona (AHC) theory~\citep{allen1976, allen1981, allen1983}. The AHC theory can be reformulated in a manner that can readily lend itself to implementation within the first-principles finite difference method: 

\begin{equation}
\label{Eq.{Enk(T)-quadratic}}
\Delta E^{\mathrm{S}}_{n\boldsymbol{k}}(T)=\frac{1}{2} \sum_{ \nu \boldsymbol{q}} \frac{1}{2 \omega_{\nu \boldsymbol{q}}} \frac{\partial^{2} E_{n\boldsymbol{k}}}{\partial \mu_{\nu \boldsymbol{q}}^{2}}\left[1+2n_{\mathrm{B}}\left(\omega_{\nu \boldsymbol{q}}, T\right)\right],
\end{equation}
where $\omega_{\nu \boldsymbol{q}}$ is the frequency of a phonon with the branch number $\nu$ and wavevector $\boldsymbol{q}$, $\mu_{\nu \boldsymbol{q}}$ is the real-valued phonon displacement (see the detailed definition in Eqs.\,\eqref{Eq.{munuq}} and \eqref{Eq.{munuq2}}), and  $n_{\mathrm{B}}(\omega_{\nu \boldsymbol{q}}, T) =[\exp(\frac{\omega_{\nu \boldsymbol{q}}}{ k_{\mathrm{B}} T})-1]^{-1}$ is the Bose-Einstein factor. It is worth noting that there is a nonzero correction to the energy band even at zero temperature with the vanishing Bose-Einstein factor, arising from the zero-point motion of the ions as a purely quantum effect. We refer readers who are interested in the rigorous mathematical derivations of the theory to Refs.\,\citep{allen1976, allen1981, allen1983, ponce2014, monserrat2018, giustino2017}, where the same results have been obtained through alternative approaches. We show the equivalence between our formulation and others in Appendix \ref{appendixa}.

\subsection{Band gap renormalized by long-wavelength phonons}
In principle, Eq.\eqref{Eq.{Enk(T)-quadratic}} captures the coupling between electrons and \textit{all}-wavelength phonons. However, in practice, explicitly accessing the $\boldsymbol{q}$-points close to the center of the Brillouin zone can become computationally prohibitive due to the inherent incompatibility of the long-wavelength limit with the Born-von-Kármán periodic boundary condition. As a result, accounting for the band gap renormalization contributed by long-wavelength phonons requires a separate treatment. 

Specifically, in the long-wavelength limit, the significance of atomic-scale interactions diminishes, and electrons can be regarded as becoming coupled with the macroscopic electric field induced by longitudinal optical phonons. This mechanism is known as the Fröhlich interaction~\cite{frohlich1950, frohlich1954}. It is especially crucial in the case of ionic crystals, where the interaction has been shown to be strong enough to significantly influence physical phenomena such as electron lifetimes and carrier mobilities~\cite{verdi2015}.

Nery and Allen derived an analytical expression to address the missing portion of the contribution associated with the long-wavelength optical phonon modes for three-dimensional (3D) materials~\cite{nery2016}. Following a similar methodology, combined with the latest results on polarons in two-dimensional (2D) systems~\cite{sio2022, sio2023}, we have derived the corresponding band gap renormalization for a 2D material:
\begin{equation}
\label{Eq.{frohlich_re_2d}}
\begin{aligned}
\Delta E^{\mathrm{L}}_{n \boldsymbol{k}}(T)=  &\int_0^{\infty}{\mathrm{d} q} \; \frac{1}{2}m^*q_0^2{d} \omega_{\mathrm{LO}}\left(\epsilon_0-\epsilon_{\infty}\right) \frac{q}{(q_0+q)^2} \\ & \left[\frac{1+n_{\mathrm{B}}\left(\omega_{\mathrm{LO}}, T\right)}{-q^2-2m^*\omega_{\mathrm{LO}}} +  \frac{n_{\mathrm{B}}\left(\omega_{\mathrm{LO}}, T\right)}{-q^2+2m^*\omega_{\mathrm{LO}}}\right]. 
\end{aligned}
\end{equation}
Here, $m^*$ is the effective mass of the band, $q_0$ is a characteristic wavevector of the polaron in 2D materials (defined in Eq.\,\eqref{Eq.{q0}} and Refs.\,\cite{sio2022, sio2023}), $d$ is the effective thickness of the 2D material, $\omega_{\mathrm{LO}}$ is the frequency of the longitudinal-optical phonon, and $\epsilon_{\infty}$ and $\epsilon_{0}$ are the high-frequency and static relative permittivities, respectively. The details of the derivation are presented in Appendix~\ref{appendixb}, where, very interestingly, we find that the dimensionality change from three to two significantly amplifies the mathematical intricacy involved in band renormalization.

\subsection{Thermal expansion}
In the context of experimentally probing the temperature dependent electron eigenenergy, a prevailing practice entails the execution of experiments under constant-pressure conditions. Therefore, it is necessary to also include the influence of thermal expansion on electron eigenenergies, which has been shown to make a comparable contribution to electron-phonon coupling in three dimensional topological materials~\cite{monserrat2016}. The temperature dependence of the electron eigenenergy induced by thermal expansion can be expressed as~\cite{allen1981, allen1983}:
\begin{equation}
\begin{aligned}
\label{Eq.{thermalexp}}
\Delta E^{\mathrm{TE}}_{n\boldsymbol{k}}(T)= \int_0^{T}{\mathrm{d} T} \; \left(\frac{\partial E_{n \boldsymbol{k}}}{\partial  V}\right)_{T_0}\left(\frac{\partial  V}{\partial T}\right)_{P_0}.
\end{aligned}
\end{equation}
Here, $(\frac{\partial E_{n k}}{\partial  V})_{T_0}$ represents the eigenenergy shift as the system volume changes at constant temperature $T_0$, and $(\frac{\partial  V}{\partial T})_{P_0}$ represents the volume change as the system temperature increases at constant pressure $P_0$.

\subsection{First-principles implementation}
To calculate band renormalization induced by short-wavelength phonons from first principles, we use Eq.\,\eqref{Eq.{Enk(T)-quadratic}} as our starting point. This choice (instead of Eq.\,\eqref{Eq.{E_nk(T)-g}}) enables us to bypass the explicit calculation of the electron-phonon coupling matrix elements. The only terms that need to be numerically evaluated are the second order derivatives of the band structure with respect to the displacement of each phonon. We implement this in the finite difference context using a three-point central formula:
\begin{equation}
\frac{\partial^{2} E_{n \boldsymbol{k}}}{\partial \mu_{\nu \boldsymbol{q}}^{2}} \simeq \frac{E_{n \boldsymbol{k}}^{\left(+\delta \mu_{\nu \boldsymbol{q}}\right)}+E_{n \boldsymbol{k}}^{\left(-\delta \mu_{\nu \boldsymbol{q}}\right)}-2 E_{n \boldsymbol{k}}^{(0)}}{\delta \mu_{\nu \boldsymbol{q}}^{2}},
\end{equation}
where $E_{n \boldsymbol{k}}^{\left(+\delta \mu_{\nu \boldsymbol{q}}\right)}$ represents the band structure calculated by incorporating a ``frozen'' phonon characterized by $\nu$ and $\boldsymbol{q}$ with a real displacement of $+\delta \mu_{\nu \boldsymbol{q}}$. In the case of degenerate band structures, the above second derivatives are averaged over all degenerate states. The BZ integration required by Eq.\,\eqref{Eq.{Enk(T)-quadratic}} is implemented by invoking the non-uniform $\boldsymbol{q}$-point sampling and nondiagonal supercell techniques~\cite{chen2022, Lloyd2015} to ensure optimal efficiency.

To calculate band renormalization induced by long-wavelength phonons, we employ Cauchy's principal value integration to evaluate Eq.\,\eqref{Eq.{frohlich_re_2d}}, circumventing the divergence arising from the singularity in the integrand when $-q^2 \pm 2m^*\omega_{LO} = 0$. We note that Ref.\,\cite{nery2016} proposed the alternative approach of invoking the adiabatic approximation to merge the two singularities into one and shifting them off the real axis by an amount $\mathrm{i}\eta$. Although we have also derived the counterpart for 2D materials following the same idea (as shown in Eq.\,\eqref{Eq.{frohlich_re_2d_ana}}), which might provide accurate results with a less dense $\boldsymbol{q}$-point grid when $\mathrm{i}\eta$ is judiciously chosen, we opt not to adopt this approach in this work to avoid introducing the \textit{ad hoc} parameter. All the parameters within Eq.\,\eqref{Eq.{frohlich_re_2d}} can be acquired through well-established first-principle methods. For instance, $m^*$ can be determined by fitting the curvature of the band, while $\epsilon_{\infty}$ and $\epsilon_{0}$ can be calculated by assessing the response of the system to an applied electric field. 

To model thermal expansion from first principles through $(\frac{\partial  V}{\partial T})_{P_0}$ in Eq.\,\eqref{Eq.{thermalexp}}, we use the quasi-harmonic approximation to incorporate the volume dependence of phonon frequencies, such that the Helmholtz free energy of the vibrating lattice can be expressed as~\citep{leibfried1961}:
\begin{equation}
\label{Eq.{free_energy_qha}}
F(V ; T)= U(V;  0) + {k_{\mathrm{B}} T}  \sum_{\nu \boldsymbol{q}} \ln \left[2\sinh \left(\frac{\omega_{\nu \boldsymbol{q}}(V)}{2 k_{\mathrm{B}} T}\right)\right],
\end{equation}
where the first term is the potential energy for the static lattice and the second term is the vibrational contribution to the free energy.  It is worth noting that the second term does not vanish at zero temperature; instead, it equals $\frac{1}{2} \sum_{\nu \boldsymbol{q}} \omega_{\nu\boldsymbol{q}}(V)$, which leads to the contribution from zero-point motion to thermal expansion. We further implement Eq.\,\eqref{Eq.{free_energy_qha}} by calculating the system of interest at a few volumes in the expansion and compression regime. At each volume, the static lattice potential energy $ U(V;  0)$ can be obtained through a lattice-constrained geometry optimisation, and the volume-dependent phonon frequency $\omega_{\nu \boldsymbol{q}}(V)$ can be obtained from the corresponding phonon calculations. By fitting the resulting Helmholtz free energy $F(V;T)$ with a polynomial function, we determine the volume corresponding to the minimum of $F(V;T)$ for a given temperature $T$, that is, the equilibrium volume at that temperature.

\section{Results and discussion}
\label{sec: iii}
\subsection{Crystal structure}

\begin{figure}
\begin{tabular}{cc}
\includegraphics[width=1.0\linewidth]{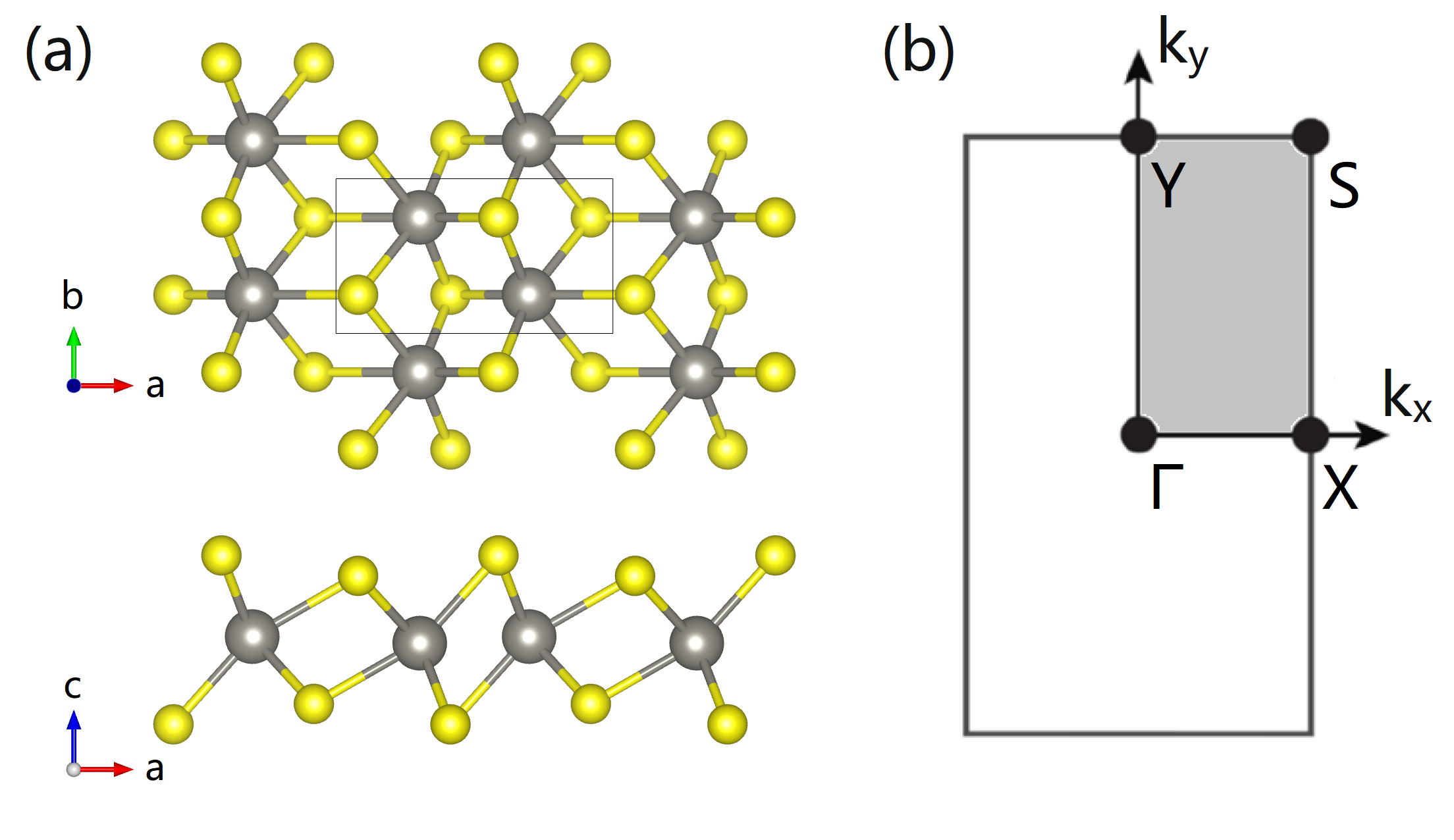} 
\end{tabular}
\caption{(a) Crystal structure of 1T$^\prime$-MX$_2$ and (b) its BZ. The gray and yellow balls represent the M=W, Mo and X=S, Se atoms respectively. The gray shaded area represents the irreducible Brillouin zone.}
\label{fig: mx2_struture}
\end{figure}
Monolayer TMDs have been confirmed to exhibit a series of stable and metastable phases, both in theory and experiments~\cite{singh2015, chou2015}. Of particular interest within this work is the 1T$^\prime$-phase, which adopts a monoclinic structure characterized by the space group $\mathsf{P}2_1/m$. Figure\,\,\ref{fig: mx2_struture}(a) visualizes the crystallographic structure from both top and side perspectives. The structure emerges as a spontaneous symmetry breaking from the 1T-phase (with space group $\mathsf{P}\bar{3}m1$) induced by the Peierls instability in which the M-M dimerization takes place thereby eliminating the degeneracy of the states near the Fermi level and lowering the energy of the system. Despite the distortion of M atoms leading to the loss of certain symmetries, the 1T$^\prime$-phase still possesses an inversion centre, a mirror plane and a two-fold screw axis. Therefore, its irreducible Brillouin zone is a quarter of the whole Brillouin zone, as depicted in Fig.~\ref{fig: mx2_struture}(b).

\subsection{Computational details}
We perform first-principles calculations at the density functional theory (DFT) level~\cite{DFT-Hohenberg-Kohn, DFT-Kohn-Sham} with the Vienna \textit{ab initio} Simulation Package ({\sc vasp})~\cite{VASP-Original-Paper}. A vacuum layer with a thickness of 20 Å is used in the calculation to avoid periodic image interactions along the direction perpendicular to the plane. The interaction between ions and valence electrons is modeled with pseudopotentials based on the projector-augmented wave~\cite{VASP-PAW-One, VASP-PAW-Two} method, where the valence electrons of the transition metals are $ns^{2}(n-1)d^{5}$ (where $n=6$ for M=W and $n=5$ for M=Mo) and the valence electrons of chalcogen are $ns^2np^4$ (where $n=4$ for X=Se and $n=3$ for X=S). The exchange-correlation functional is treated in the generalized gradient approximation parametrized by Perdew-Burke-Ernzerhoff (PBE)~\cite{PBE-exchange-correlation}. An energy cut-off of $500$\,eV for the plane-wave expansion and a $\Gamma$-centered $\boldsymbol{k}$-point grid of size $5\times9\times1$ for Brillouin zone integration are adopted in the calculations. The spin-orbit coupling is included in the calculations \textit{via} a perturbation to the scalar relativistic Hamiltonian~\cite{koelling1977}.

\subsection{Electronic structures}

\begin{table}
\begin{ruledtabular}
\begin{tabular}{cccccc}
   &  & WS$_2$ & WSe$_2$ & MoS$_2$ & MoSe$_2$ \\
\hline
  & static & $183$\,meV & $693$\,meV & $531$\,meV & $713$\,meV  \\
  \multirow{2}{*}{$\Delta E^\mathrm{S}$} & 0\,K & $+17$\,meV & $+5$\,meV & $+4$\,meV & 	$+8$\,meV  \\
  & 300\,K & $+38$\,meV & $+17$\,meV & $+17$\,meV & $+21$\,meV  \\
  \multirow{2}{*}{$\Delta E^\mathrm{L}$} & 0\,K & $-8$\,meV & $-5$\,meV & $-2$\,meV & $-5$\,meV \\
  & 300\,K & $-8$\,meV & $-5$\,meV & $-1$\,meV & $-5$\,meV \\
  \multirow{2}{*}{$\Delta E^\mathrm{TE}$} & 0\,K & $-4$\,meV & $-3$\,meV & $-2$\,meV & $-0.4$\,meV \\
  & 300\,K & $-17$\,meV & $-16$\,meV & $-13$\,meV & $-1$\,meV \\
\end{tabular}
\end{ruledtabular}
\caption{Static and renormalized band gap of WS$_2$, WSe$_2$, MoS$_2$, and MoSe$_2$ at $T=0$\,K and 300\,K. The total renormalization is equal to the sum of the renormalization induced by short- and long-wavelength phonons, $\Delta E^{\mathrm{S}}(T)$ and $\Delta E^{\mathrm{L}}(T)$, and the thermal expansion, $\Delta E^{\mathrm{TE}}(T)$.}
\label{tab:bandgap}
\end{table}

\begin{figure}
\centering
\includegraphics[width=1.01\linewidth]{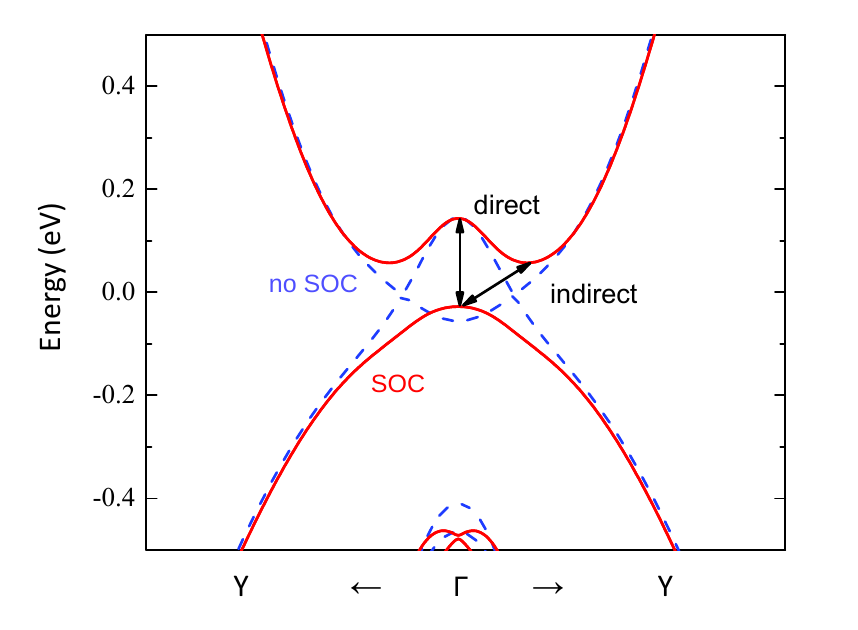} 
\caption{Band structures of WS$_2$ with and without spin-orbit coupling (SOC) in the vicinity of the $\Gamma$ point. The direct and indirect gaps are highlighted, where the former corresponds to the inverted band gap in the context of topological insulators.}
\label{fig:band_ws2}
\end{figure}

\begin{figure*}
\,
\includegraphics[width=0.32\linewidth]{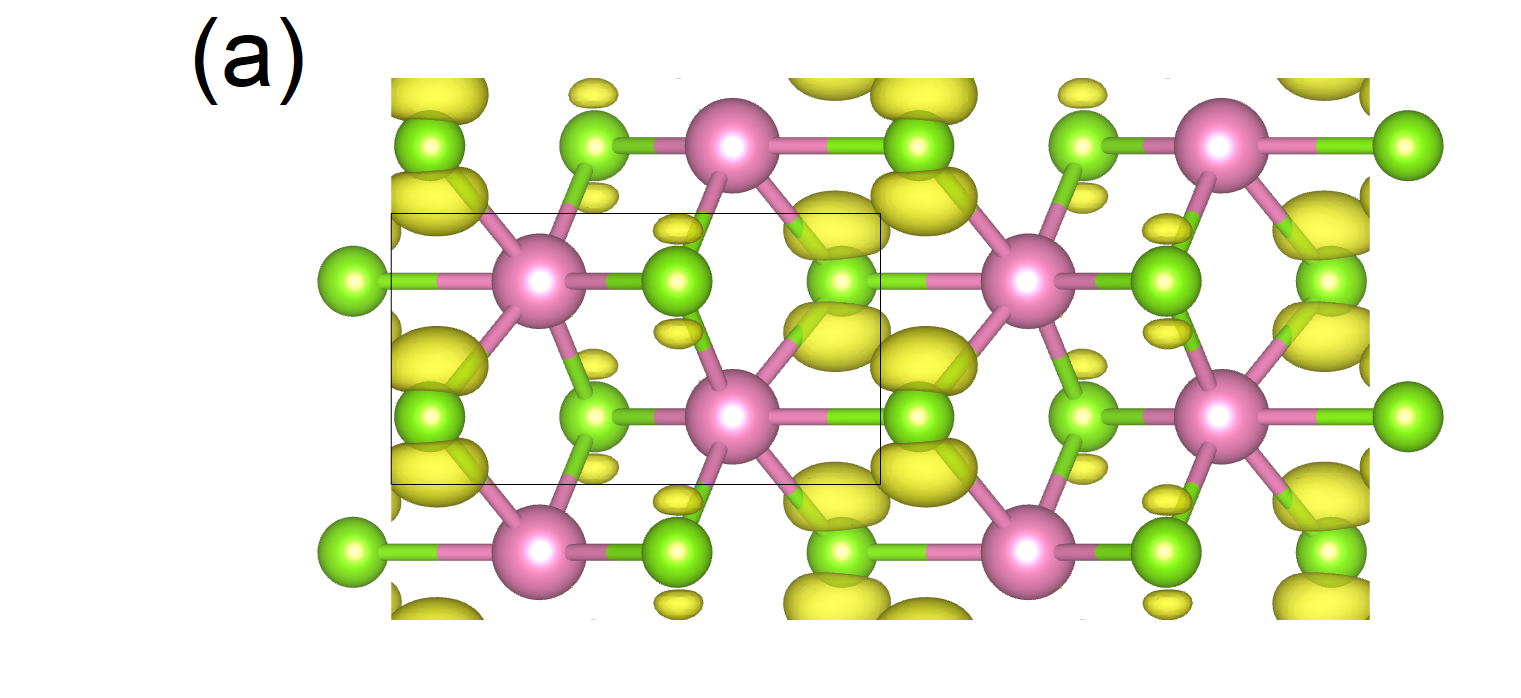}
\includegraphics[width=0.32\linewidth]{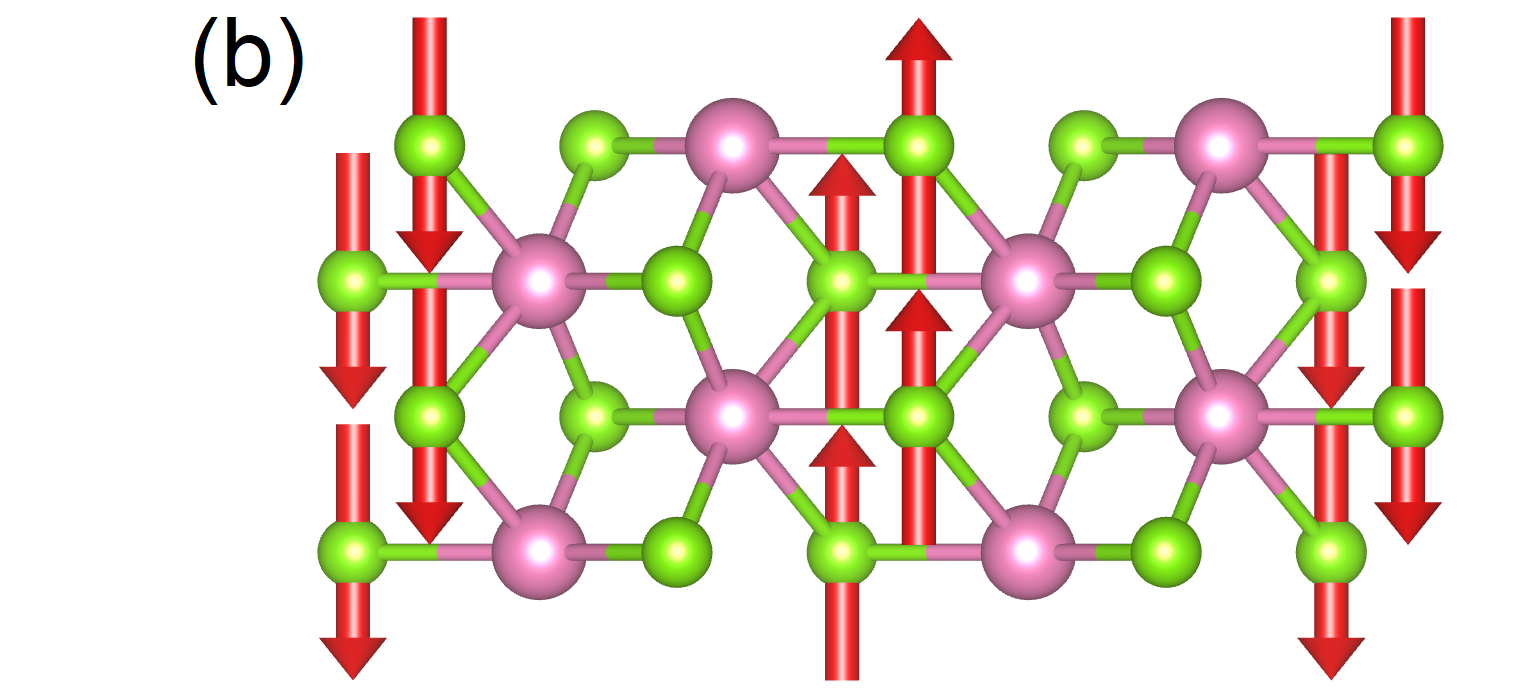} 
\includegraphics[width=0.32\linewidth]{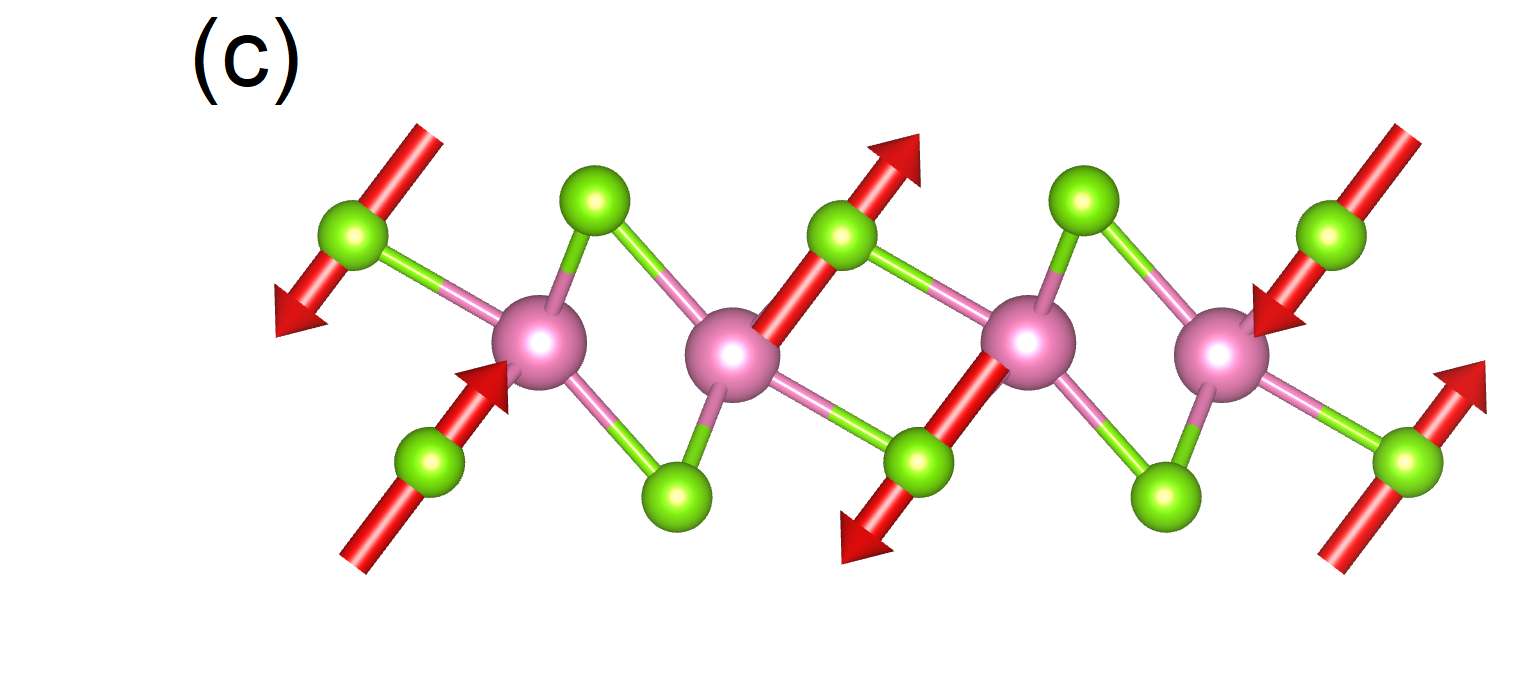} 
\newline
\newline
\includegraphics[width=0.32\linewidth]{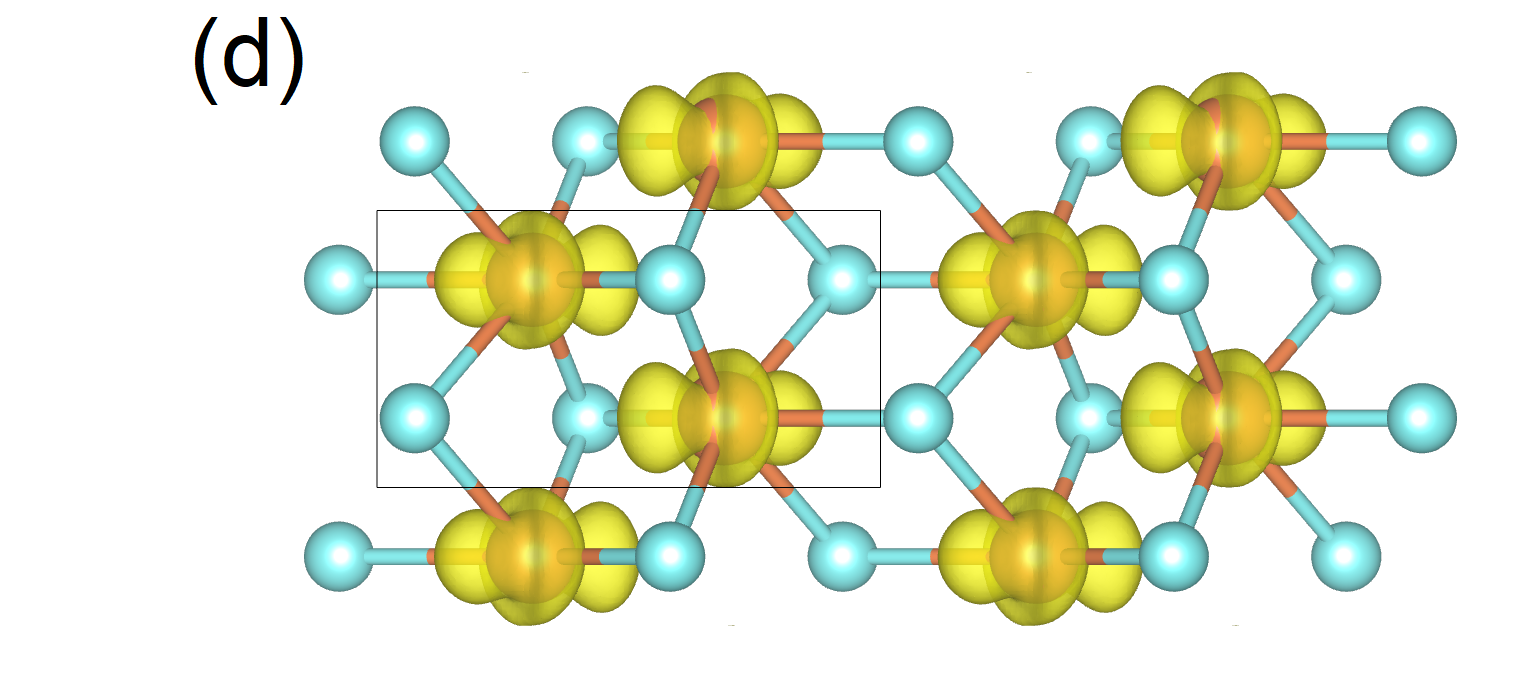}
\includegraphics[width=0.32\linewidth]{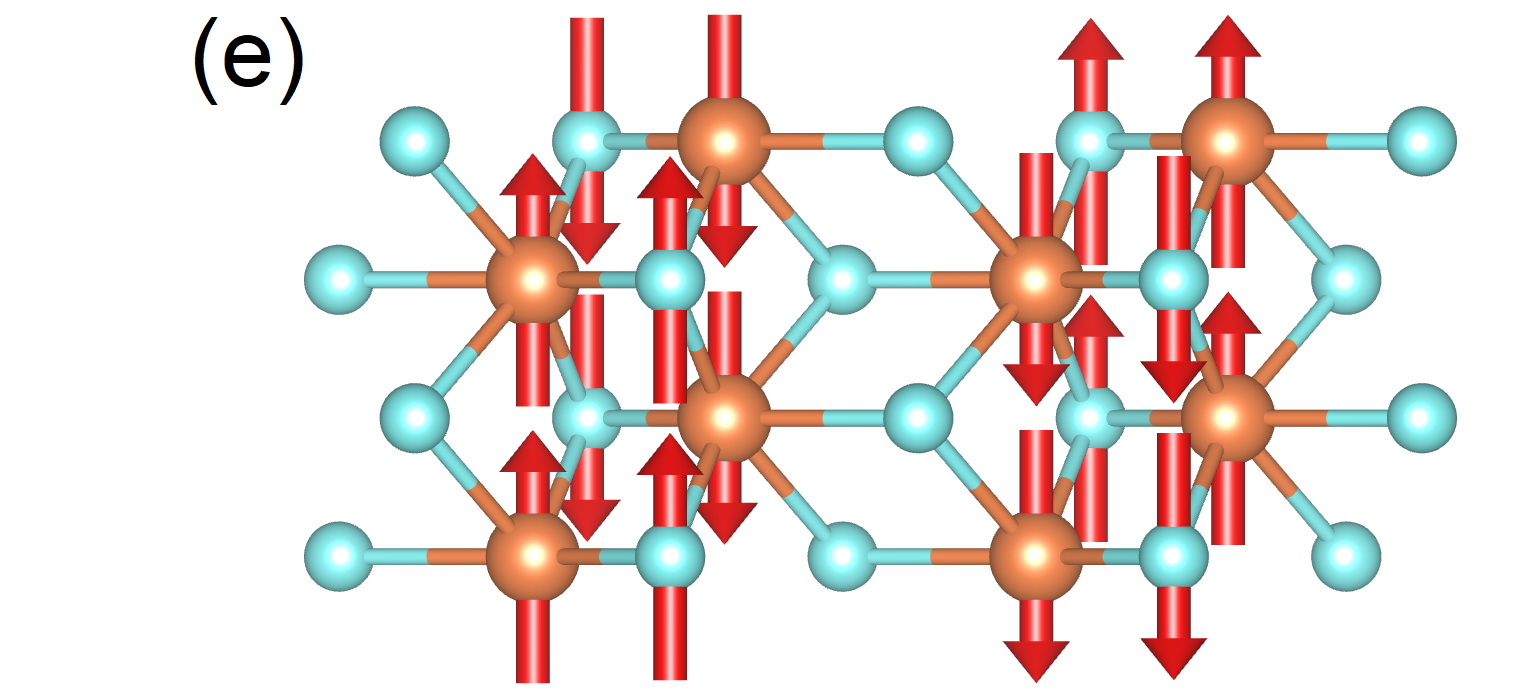} 
\includegraphics[width=0.32\linewidth]{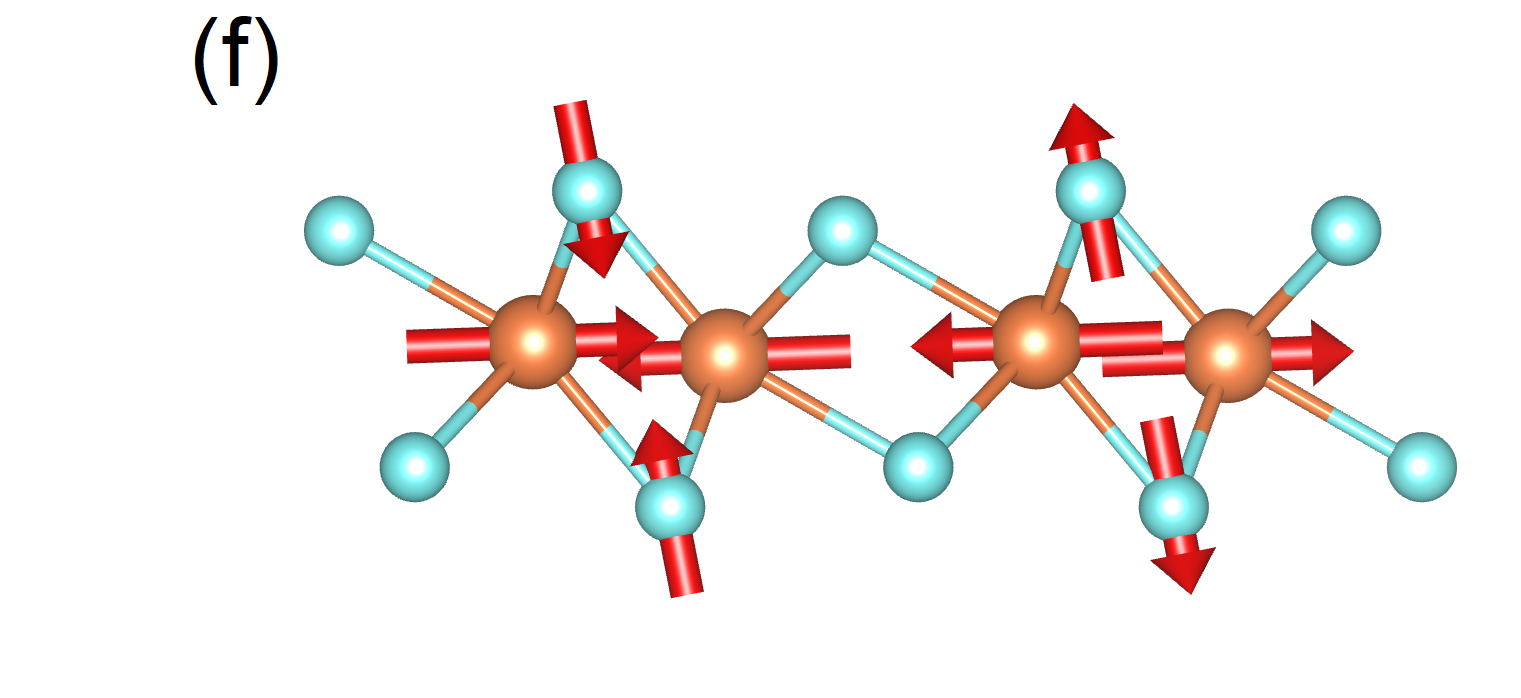} 
\caption{Hole density of the state associated with the conduction band minimum at the $\Gamma$ point, alongside two vibration modes at the X point that exhibit strong coupling with it. Panels (a-c) correspond to MoS$_2$, WS$_2$ and WSe$_2$, while panels (d-f) correspond to MoSe$_2$. The red arrows indicate the vibration of the corresponding atoms.}
\label{fig:mx2_density}
\end{figure*}

It has been confirmed both in theory and experiments that the 1T$^\prime$-MX$_2$ family are topological insulators~\cite{qian2014, ugeda2018, xu2018, chen2018, jelver2019, cayssol2021, xu2023}. In a broader context, the band structure of an inversion symmetric topological insulator is generally characterized by two band gaps, the indirect band gap, which corresponds to the minimum energy to excite an electron in the material bulk, and the direct band gap, where the band inversion takes places. Using WS$_2$ as a representative example, we show a typical band structure of a topological insulator in Fig.\,\ref{fig:band_ws2}, where the band inversion induced by spin-orbit coupling occurs at the $\Gamma$ point and leads to a double-well-shaped dispersion for the conduction band.

Following Ref.\,\cite{qian2014}, we refer to the direct band gap at the $\Gamma$ point as the inverted band gap, which reflects the band inversion strength of MX$_2$. Table\,\,\ref{tab:bandgap} summarizes the values of the inverted band gaps for WS$_2$, WSe$_2$, MoS$_2$, and MoSe$_2$ at the static DFT level, where MoSe$_2$ exhibits the largest gap of 713\,meV while WS$_2$ exhibits the smallest one of 183\,meV. The results agree with previous first-principles studies~\cite{qian2014, das2020}. 


The atomic orbital projection analysis of the inverted band gap point further reveals that the non-trivial topology of 1T$^\prime$-MX$_2$ can be understood from the $p-d$ band inversion picture. We find that across all four compounds, the conduction band inversion is primarily influenced by hybridization of metal $d_{xz}$ and $d_{yz}$ states. In the cases of MoS$_2$, WS$_2$, and WSe$_2$, their inverted conduction bands exhibit a similar pattern, predominantly stemming from contributions of the chalcogenide $p_y$ orbitals. On the contrary, MoSe$_2$ distinguishes itself from them, displaying an inverted conduction band dominated by a mixture of Mo $d_{x^2-y^2}$ and $d_{z^2}$ states. This state manifests a distinctly different orientation compared to the chalcogenide $p_y$ orbitals, as shown in Figs.~\ref{fig:mx2_density}(a) and (d).

\subsection{Band gap renormalization induced by electron-phonon coupling}

To effectuate an electronic structure phase transition from a topologically nontrivial state (\ie 1T$^\prime$-MX$_2$ at 0\,K) to a topologically trivial state (if exists), it is imperative for the inverted band gap at the $\Gamma$ point to undergo closure.  Therefore, in subsequent sections, we focus on the temperature effects on this inverted band gap. It is also worth noting that, as one of the most important manifestations of topology, the topologically protected metallic edge states which carry opposite spin polarizations are required to intersect at the $\Gamma$ point because of time-reversal symmetry. The presence of a finite gap guarantees the insulating nature of the bulk and ensures that electronic transport can occur only through the topologically protected metallic edge states.

\subsubsection{Short-wavelength phonon-induced band gap renormalization}

\begin{figure}
\centering
\includegraphics[width=1.01\linewidth]{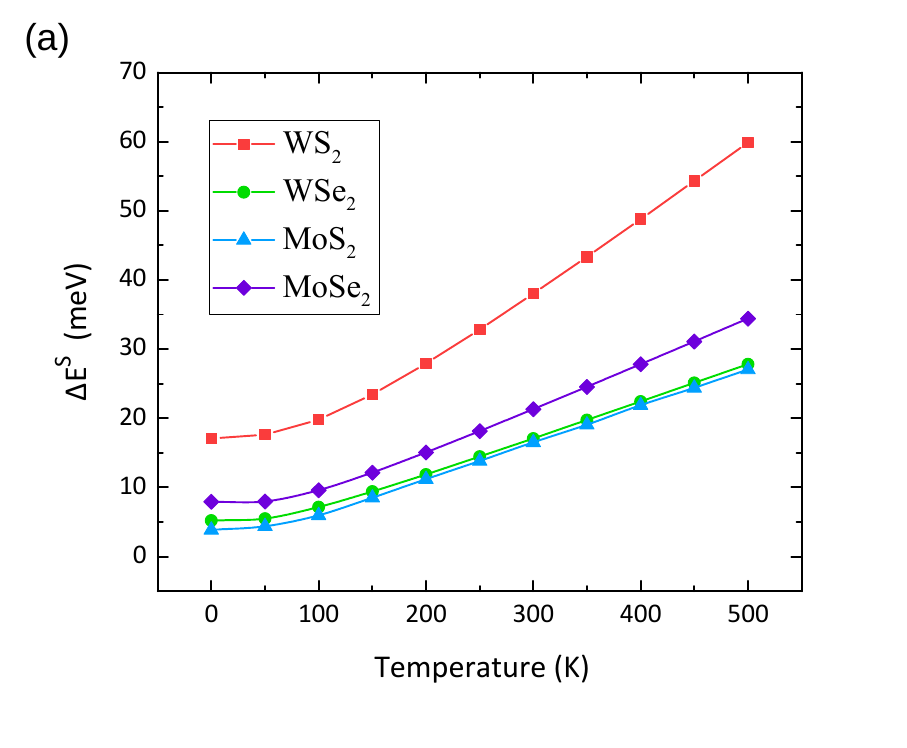} 
\includegraphics[width=1.01\linewidth]{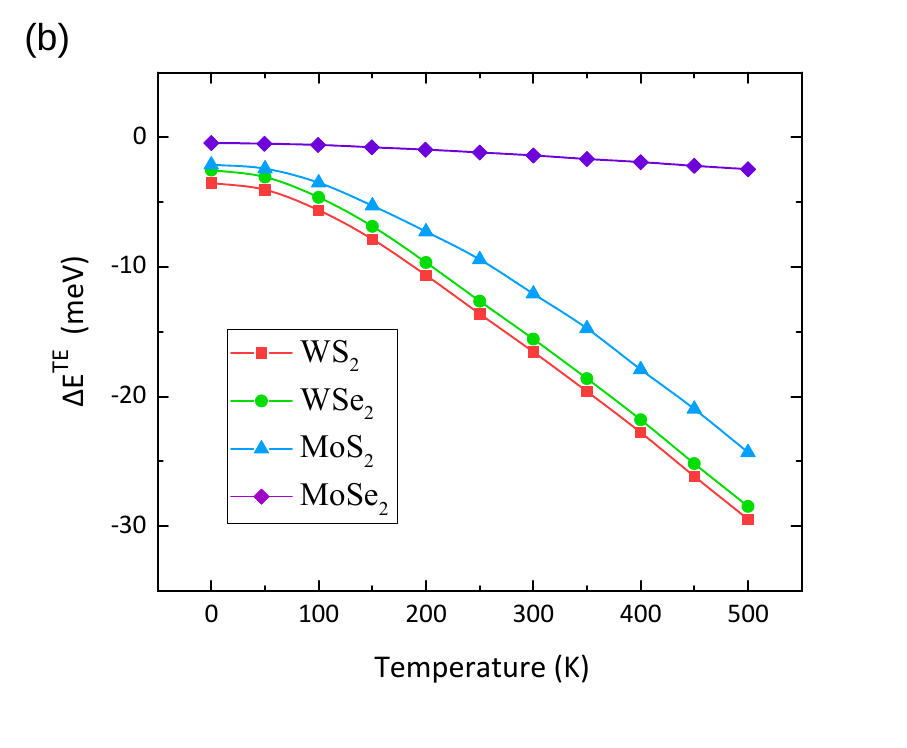} 
\includegraphics[width=1.01\linewidth]{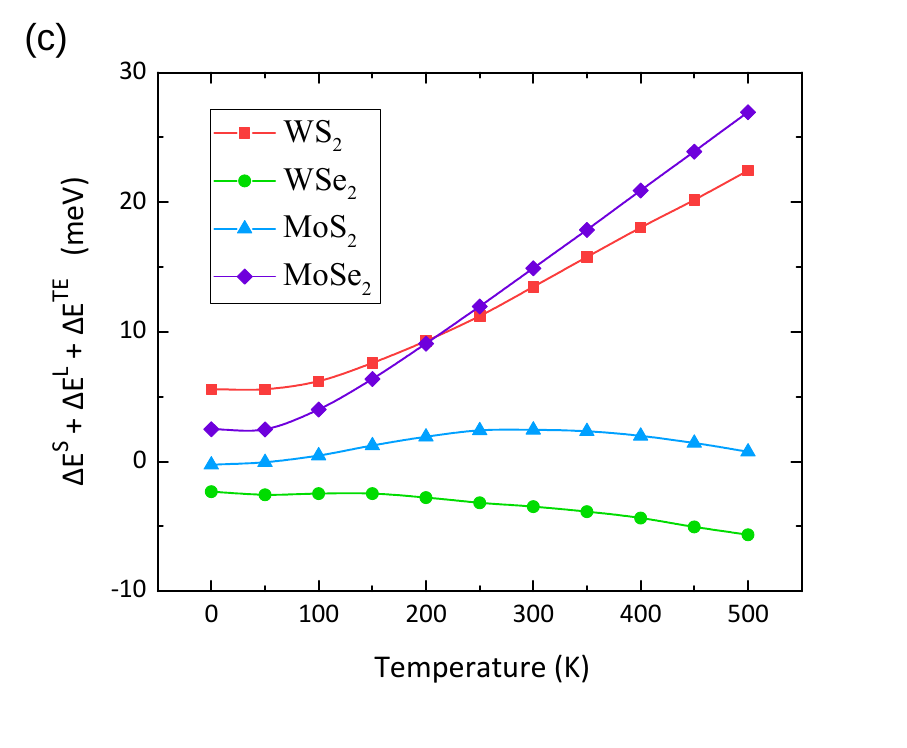} 
\caption{Band gap renormalization as a function of temperature for WS$_2$, WSe$_2$, MoS$_2$ and MoSe$_2$, where (a) and (b) show the contribution from short-wavelength phonons, $\Delta E^{\mathrm{S}}(T)$, and thermal expansion, $\Delta E^{\mathrm{TE}}(T)$ respectively, and (c) shows the overall effect including both electron-phonon coupling and thermal expansion.}
\label{fig:band_vs_T}
\end{figure}

\begin{figure*}
\includegraphics[width=0.488\linewidth]{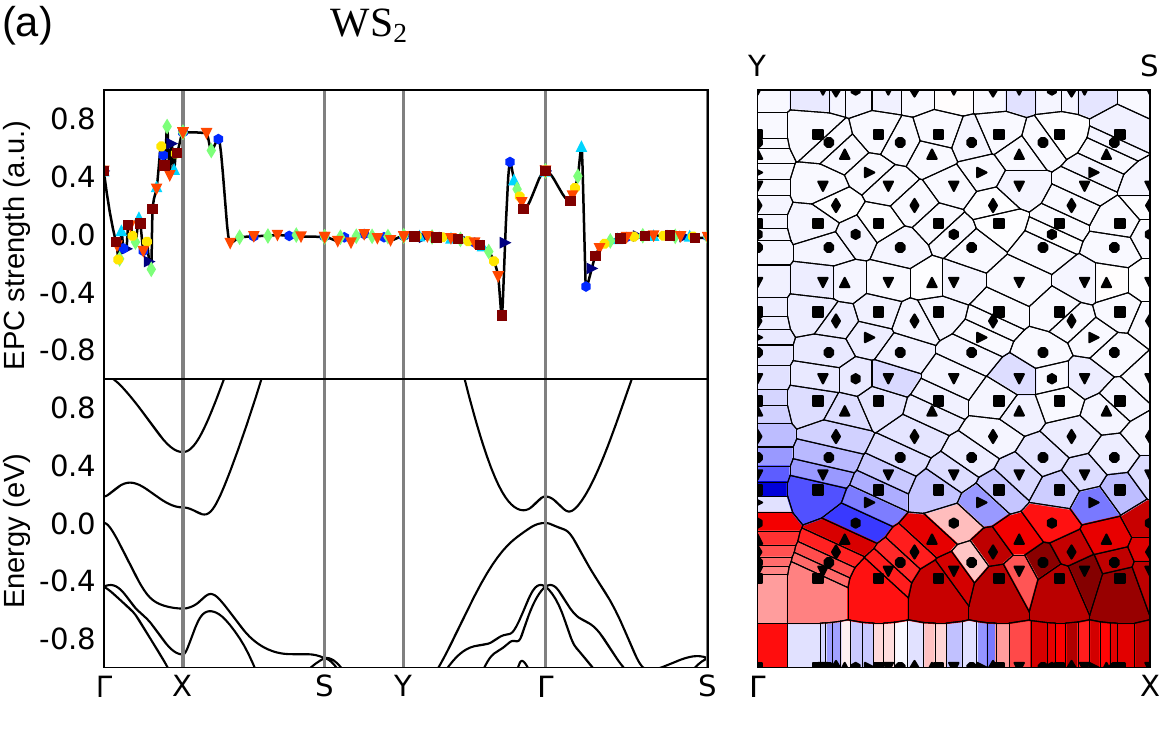} \quad
\includegraphics[width=0.488\linewidth]{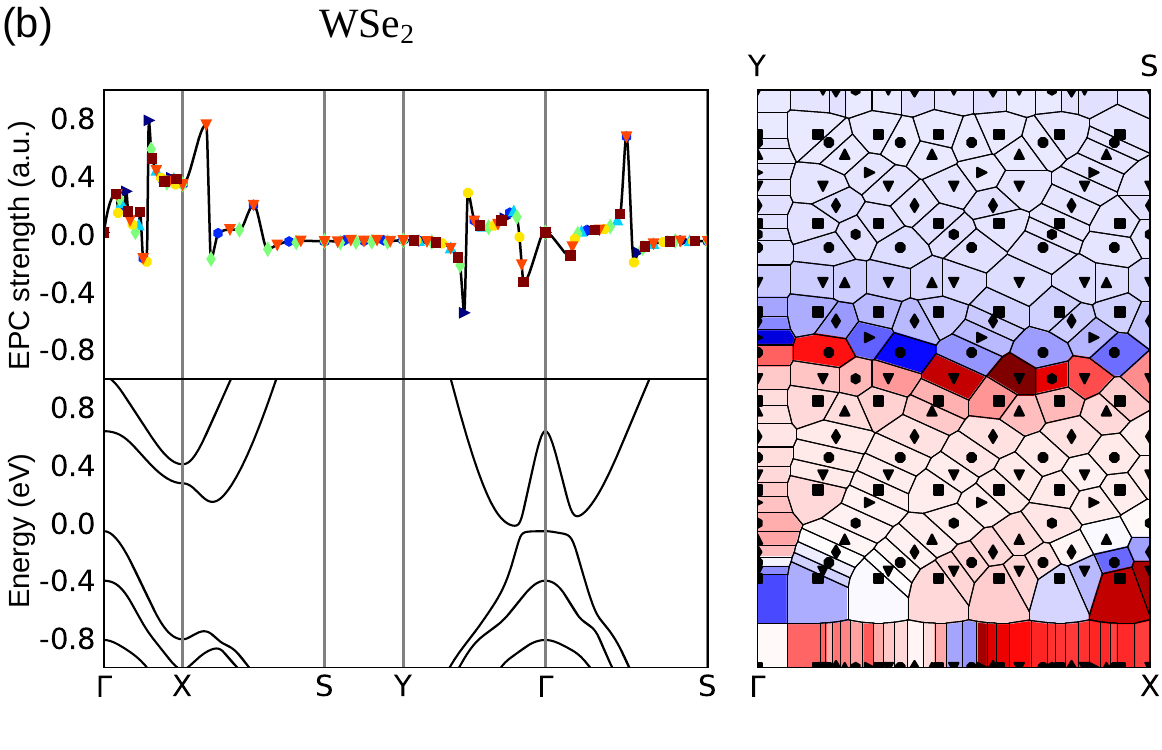} \\
\includegraphics[width=0.488\linewidth]{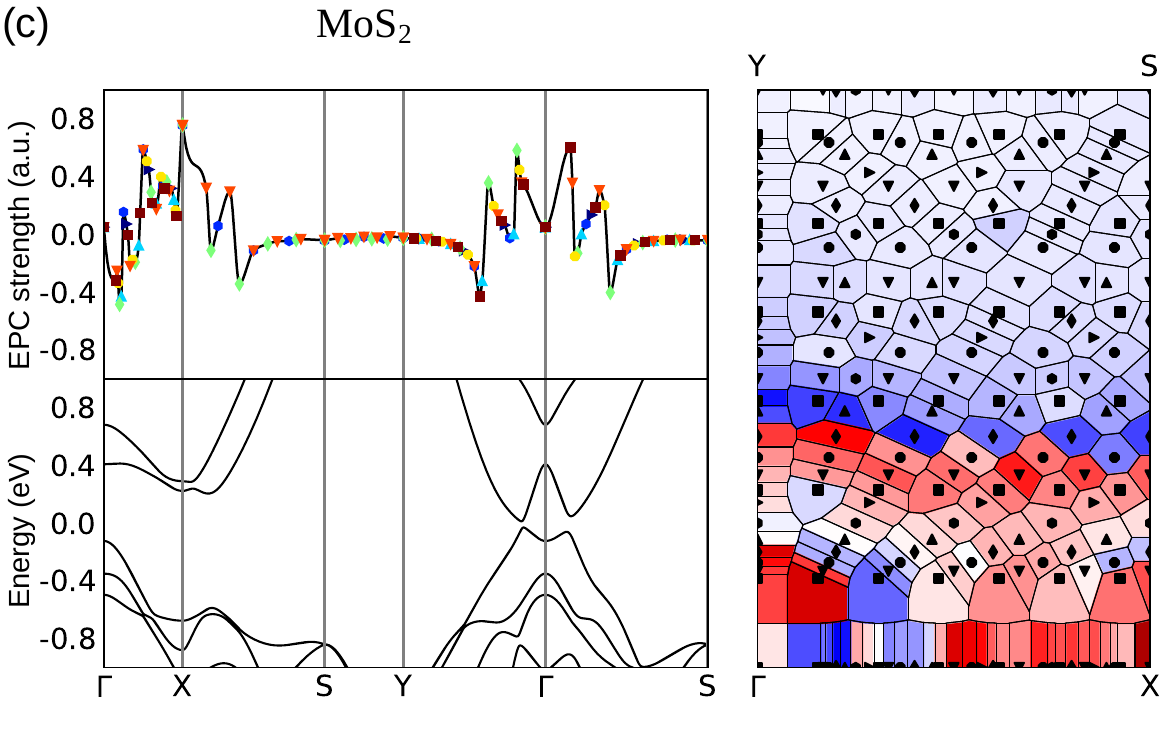} \quad
\includegraphics[width=0.488\linewidth]{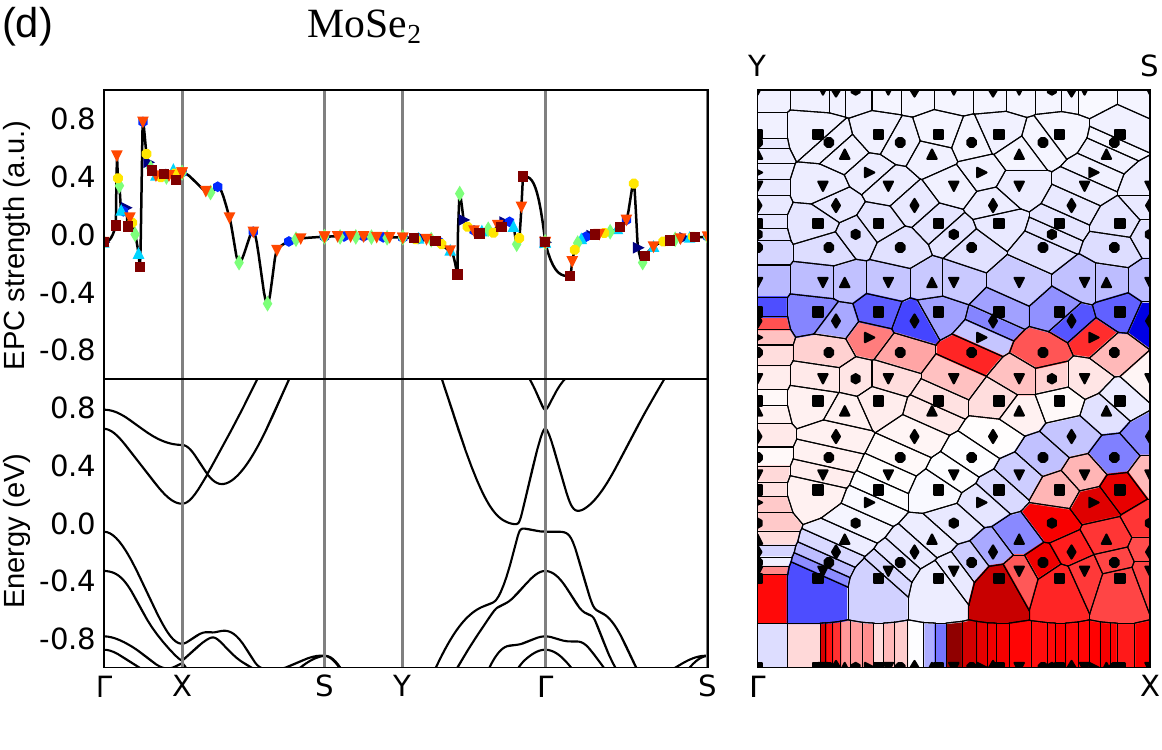} 
\caption{Electron-phonon coupling strength (left top panel) and band structure (left bottom panel) for (a) WS$_2$, (b) WSe$_2$, (c) MoS$_2$, and (d) MoSe$_2$ along a high symmetry path in the irreducible Brillouin zone. To calculate the band gap renormalization, the irreducible Brillouin zone is sampled by a non-uniform Farey grid as shown in the right panel, where red and blue represent positive and negative electron-phonon coupling strengths respectively, and the color depth represents the absolute value of the strength.}
\label{fig:el-ph-strength}
\end{figure*}

To obtain the phonon-induced band gap renormalization, we first calculate the phonon spectra for the four compounds thereby determining the concrete form of real-valued operators $\frac{\partial}{\partial \mu_{\nu \boldsymbol{q}}}$ for them. The obtained phonon spectra also allow us to theoretically confirm the dynamical stability of MX$_2$ at ambient pressure as there are no imaginary vibrational modes (see Fig.\,S1 in Supplementary Material (SM) for details).  

Figure\,\,\ref{fig:band_vs_T}(a) shows the band gap corrections $\Delta E^\mathrm{S}$ as a function of temperature, while the specific values at both absolute zero and room temperature ($T=300$ K) are presented in Table\,\,\ref{tab:bandgap}. In contrast to the behavior typically observed in most semiconductors, where the band gap diminishes with rising temperature, 1T$^\prime$-MX$_2$  exhibits an opposite trend. Across the four TMDs investigated, regardless of the distinct chemical compositions, a consistent positive correlation between the band gap and temperature is evident. It is particularly pronounced at temperatures exceeding 100\,K, showing an asymptotic linear dependence of the gap change with temperature. Below this temperature, only the low-frequency vibrational modes are excited and the phonon-induced band gap renormalization remains relatively constant. For WSe$_2$, MoS$_2$, and MoSe$_2$, we find that they exhibit comparable band gap corrections, around 20\,meV at room temperature. In comparison, WS$_2$ manifests stronger electron-phonon coupling, resulting in a band gap renormalization of $38$\,meV, which is approximately double the values observed in the other three TMDs. This is a substantial correction, whose magnitude accounts for nearly 21\% of the static band gap size in WS$_2$.

To have better understanding of the microscopic origin of the phonon-induced band gap renormalization, we define the electron-phonon coupling strength in the temperature dependent band structure context as
\begin{equation}
S^{\mathrm{el\text{-}ph}}_{\nu \boldsymbol{q}} = \frac{\partial^2 E_{\mathrm{gap}}}{\partial \mu_{\nu \boldsymbol{q}}^2},
\end{equation}
followed by the $\boldsymbol{q}$-resolved electron-phonon coupling strength defined as
\begin{equation}
S^{\mathrm{el\text{-}ph}}(\boldsymbol{q}) = \sum_{\nu} S^{\mathrm{el\text{-}ph}}_{\nu \boldsymbol{q}}.
\end{equation}

The left bottom panels of Figs.~\ref{fig:el-ph-strength} (a-d) depict the band structures for the four compounds along the high symmetry path, from which it can be seen that their conduction band minima are all located on the $\Gamma-Y$ path. For WS$_2$ and MoSe$_2$, they exhibit a (negative) parabolic and quartic shaped valence band around the $\Gamma$-point respectively, while for WSe$_2$ and MoS$_2$, their valence band manifest a double-peak shape, resulting in the valence band maximum shifting close to the conduction band maximum on the $\Gamma-Y$ path. 

The left top and right panels of Figs.\,\ref{fig:el-ph-strength} (a-d) show the $\boldsymbol{q}$-resolved electron-phonon coupling strength along the high-symmetry path and on the whole irreducible Brillouin zone, respectively. Within the upper half-plane of the irreducible Brillouin zone, the electron-phonon coupling strength is nearly negligible. The principal contributions to the band renormalization are concentrated in proximity to the $\Gamma$ and $X$ points. Focusing on phonons at the X point, which corresponds to vibrational modes exhibiting a wavelength twice that of the original unit cell, we visualize in Figs.\,\ref{fig:mx2_density}(b-c) and (e-f) the two modes that most strongly couple to the inverted band gap. Interestingly, these modes are highly localised along certain directions. For MoS$_2$, WS$_2$ and WSe$_2$, the dominant modes are very similar, all exclusively associated with chalcogenide atoms. One mode involves vibrations along the $\boldsymbol{b}$-direction, inducing alternating shear deformation, while the other involves vibrations along the $\boldsymbol{a}$- and $\boldsymbol{c}$-directions, asymmetrically stretching chalcogenide atoms. On the other hand, MoSe$_2$ shows a very different behaviour, with the strong coupling modes governed by atoms that remain static in the cases of MoS$_2$, WS$_2$, and WSe$_2$. This difference can be attributed to its distinct orbital component. As mentioned earlier, MoS$_2$, WS$_2$, and WSe$_2$ display a conduction band at the $\Gamma$ point that mainly arises from chalcogenide $p_y$ orbitals, whereas in MoSe$_2$, it originates from Mo $d_{x^2-y^2}$ and $d_{z^2}$ orbitals. The modes we observe here can bring about a notable alteration in hole density, consequently renormalizing the inverted band gap.

Furthermore, a noteworthy finding is that within WS$_2$ alone, the inverted band gap shows a nonzero coupling to phonons at the $\Gamma$ point. The 8th phonon branch with a flat dispersion ($\omega_{\nu\boldsymbol{q}} \approx$ 28\,meV) contributes nearly 73\% electron-phonon coupling strength. This unique $\Gamma$ point contribution drives a larger band gap renormalization at room temperature in WS$_2$ compared to that of the other three compounds in the TMD family. 

Finally, it is also worth noting that the electron-phonon coupling strength curves diverge at some $\boldsymbol{k}$-points. This divergence manifests as sharp peaks in the left top panel of Figs.\,\ref{fig:el-ph-strength}(a-d) and visually evident red-blue dividing lines in the right panel. This feature arises from the double-peak and double-well characteristics of the band structure through the divergence of the electron-phonon coupling strength on the isoenergetic surface. This phenomenon can be rationalized using a perturbation theory framework and has been discussed in the case of bismuthene in Ref.\,\cite{chen2022}. The computational cost associated with treating the singularity in the electron-phonon coupling strength prevents the calculation of band gap renormalization using the traditional uniform $\boldsymbol{q}$-point grid. Therefore, here we have used a Farey grid of order 13, sampling 696 $\boldsymbol{q}$-points in the Brillouin zone, corresponding to 12528 modes, to ensure convergence (see Fig.\,S2 in SM for the detailed convergence tests of band gap renormalization).

\subsubsection{Long-wavelength phonon-induced band gap renormalization}
\begin{table}
\begin{tabular}{ccccc}
\hline\hline
 & WS$_2$ & WSe$_2$ & MoS$_2$ & MoSe$_2$ \\
\hline
\multirow{1}{*}{$\omega_{\mathrm{LO}}$}    & $1.26\times 10^{-3}$ & $9.25\times 10^{-4}$ & $1.26\times 10^{-3}$ & $9.13\times 10^{-4}$  \\
$\epsilon_0$  & $20.2$ & $15.2$ & $18.9$ & $20.9$  \\
$\epsilon_\infty$   & $19.4$ & $14.8$ & $18.6$ & $20.5$  \\
$m_{\mathrm{c}}^*$ &  $-0.553$ & $-0.099$ & $-0.127$ & $-0.076$  \\
$m_{\mathrm{v}}^*$ &  $-0.622$ & $-2.034$ & $0.798$ & $27.781$  \\
$d$  & $11.5$ & $12.2$ & $11.5$ & $12.2$\\
$q_0$  & $8.97\times 10^{-3}$ & $1.11\times 10^{-2}$ &  $9.36\times 10^{-3}$  &  1$8.0\times 10^{-3}$   \\
$\Delta E^\mathrm{L}(0\mathrm{K})$  & $-8$\,meV  & $-5$\,meV & $-2$\,meV & $-5$\,meV \\
$\Delta E^\mathrm{L}(300\mathrm{K})$  & $-8$\,meV & $-5$\,meV & $-1$\,meV & $-5$\,meV \\
\hline\hline
\end{tabular}
\caption{Material-specific parameters to evaluate the band gap renormalization induced by long-wavelength phonons for WS$_2$, WSe$_2$, MoS$_2$, and MoSe$_2$. Except for $\Delta E^\mathrm{L}$ which is in units of meV, the remaining quantities are reported in atomic units, that is, $\omega_{\mathrm{LO}}$ in units of Hatree, $m_{\mathrm{c}}^*$ and $m_{\mathrm{v}}^*$  in units of the electron mass $m_\mathrm{e}$, $d$  in units of the Bohr radius $a_0$, and $q_0$ in units of $1/a_0$. }
\label{tab:polarons}
\end{table}

Given that monolayer 1T$^\prime$-MX$_2$ possesses polar chemical bonds between M$^{4+}$ and X$^{2-}$ ions, it is worth examining the band renormalization contributed by long-wavelength phonons (\ie Fröhlich coupling). Table\,\,\ref{tab:polarons} summarizes the essential quantities characterizing the Fröhlich interaction and the corresponding band renormalization values calculated at both 0\,K and 300\,K. We find that the magnitude of the renormalization closely resembles that observed in the case induced by short-wavelength phonons at 0\,K. Nevertheless, it is worth noting that the polaron-induced band renormalization exhibits minimal sensitivity to temperature fluctuations. This remarkable temperature independence can be attributed to the very small ionic contribution to the dielectric constant,  $\epsilon_0 - \epsilon_{\infty} \approx 0.5$, across all four considered 2D TMDs. In contrast, HgTe in CdTe/HgTe/CdTe quantum wells (as another realization of topological insulators) have $\epsilon_0 - \epsilon_{\infty}=6$~\cite{saha2014}, and typical ionic compounds have even larger values~\cite{lambrecht2017}.

\subsection{Thermal expansion}
Now we consider the band gap renormalisation induced by thermal expansion. Due to the 2D materials nature, all of 1T$^\prime$-MX$_2$ have weak thermal expansion. The degree of thermal expansion is proportional to the atomic weight, where MoS$_2$ has the largest thermal expansion, but still does not exceed 0.7\% (see Fig.\,S3 in SM for details).  

Figure\,\,\ref{fig:band_vs_T}(b) shows the band gap corrections $\Delta E^\mathrm{TE}$ as a function of temperature, while the specific values at both absolute zero and room temperature ($T=300$ K) are presented in Table\,\,\ref{tab:bandgap}. Overall, the inverted band gaps of all four TMDs decrease as temperature increases. This behavior is expected to some extent, given that thermal expansion drives materials toward their atomic limits, whilst at the same time resulting in a topologically trivial band structure. However, it is worth noting that the band structures of WS$_2$, MoS$_2$, and WSe$_2$ exhibit significant sensitivity to temperature, while MoSe$_2$ displays limited temperature dependence.

It is important to highlight that our QHA model is exclusively focused on isotropic thermal expansion. Nevertheless, the established dependence of the band gap in MX$_2$ under anisotropic strain~\cite{lin2017_a} implies that the inclusion of anisotropic thermal expansion would not qualitatively alter our conclusions.

\subsection{Overall temperature dependence}

In the end, we investigate the overall temperature dependence of the inverted band gap. Figure \,\,\ref{fig:band_vs_T}(c) shows total band gap corrections as a function of temperature, taking into account the contributions from both electron-phonon coupling and thermal expansion. We find that although WS$_2$ exhibits the most pronounced electron-phonon coupling, this effect is tempered by substantial thermal expansion, thereby attenuating the band gap renormalization to some extent. In the case of MoS$_2$, the band gap renormalization reaches a maximum of 3\,meV at 300\,K. Beyond this point, thermal expansion gains prominence, leading to a reduction in the band gap, \ie the temperature dependence becomes non-monotonic. As for WSe$_2$, the zero-point motion is outweighed by thermal expansion, leading to a slight decrease in the band gap at 0\,K. In comparison, MoSe$_2$ stands out with the most substantial renormalized band gap when the temperature exceeds 200\,K. At room temperature, its band gap renormalization reaches around 15\,meV.

\section{Inverse Varshni effect driven by band inversion}
\label{sec: iv}
Our observations in the four monolayer TMDs show that there seems to be no general trend for the sign of the correction to the band gap in topological insulators: temperature can either promote or suppress the topological phase. This complexity arises from the multiple ways in which temperature can exert its effects, and the competition between these effects is highly contingent upon the very details of the system.

Nevertheless, valuable insights can be gleaned from a very simple model in which electron-phonon coupling itself is indeed conducive to promoting the topological phase when significant band inversion occurs. To illustrate this, let us first consider a normal insulator characterized by parabolic valence and conduction bands at the $\Gamma$ point where $m_\mathrm{c}>0$ and $m_\mathrm{v}<0$. The band extrema are simply coupled to all other states by a dispersionless phonon with frequency $\omega_0$. Assuming that the Debye-Waller term is considerably smaller than the Fan-Migdal term (see the definitions in Appendix~\ref{appendixa}), the predominant correction to the band gap occurs near the $\Gamma$ point where $E_{n\Gamma}^{(0)} -E_{m \Gamma+\boldsymbol{q}}^{(0)} = \pm\frac{\boldsymbol{q}^2}{2|m^*|}$ (see Eq.\,\eqref{Eq.{E_nk(T)-g}}). The plus/minus sign is for the valence/conduction band, and $m^*$ describes the absolute effective masses. As a result, the temperature dependence of the band gap takes the form:

\begin{equation}
E_{\mathrm{gap}}(T)= \begin{dcases} \frac{-C \mathrm{e}^{\omega_0/(k_\mathrm{B}T)}}{\mathrm{e}^{\omega_0/(k_\mathrm{B}T)}-1} & \text { for 3D } \\ \; \frac{A-B \mathrm{e}^{\omega_0/(k_\mathrm{B}T)}}{\mathrm{e}^{\omega_0/(k_\mathrm{B}T)}-1} & \text { for 2D }
\end{dcases}, 
\end{equation}
where $C>0$ and $B>A>0$. The negative sign arises from the opposite curvatures of the valence and conduction bands. This simplified model affords insights into the so-called ``Varshni effect''~\cite{varshni1967} observed in semiconductor physics: the reduction in the energy gap of semiconductors as a function of temperature, a phenomenon observed in the vast majority of insulators.

Interestingly, when applying this same simplified model to an insulator with an inverted band gap, characterized by $m_\mathrm{c}<0$ and $m_\mathrm{v}>0$, we find:
\begin{equation}
E_{\mathrm{gap}}(T)= \begin{dcases} \frac{C}{\mathrm{e}^{\omega_0/(k_\mathrm{B}T)}-1} & \text { for 3D } \\ \; \frac{B-A \mathrm{e}^{\omega_0/(k_\mathrm{B}T)}}{\mathrm{e}^{\omega_0/(k_\mathrm{B}T)}-1} & \text { for 2D }
\end{dcases}, 
\end{equation}
from which one can anticipate an ``inverse Varshni effect''. This is particularly relevant because topological insulators often exhibit band inversion with the above band curvature characteristics.

The same argument also holds to the band renormalization induced by Fröhlich coupling (see Eq.\,\eqref{frohlich_re_2d_ad}), where the sign of the effective mass of the electrons and holes determines the sign of the renormalization. In a more general manner, one can interpret $g_{m n \nu}(\boldsymbol{k}, \boldsymbol{q})$ in Eq.\,\eqref{Eq.{E_nk(T)-g}} as  $g_{m n \nu}(\boldsymbol{k}, \boldsymbol{q})=g^{\mathrm{S}}_{m n \nu}(\boldsymbol{k}, \boldsymbol{q}) + g^{\mathrm{L}}_{m n \nu}(\boldsymbol{k}, \boldsymbol{q})$, 
encompassing both short- and long-wavelength contributions to the electron-phonon coupling matrix element~\cite{verdi2015}. Thus, we conclude that electron-phonon coupling should broadly promote the topology of systems with a significant band inversion feature.

This insight suggests that the inverse Varshni effect should be more prevalent in topological insulators than in normal insulators. We also note that Garate and Saha reached the same conclusion by considering the renormalization of Dirac mass (instead of the band gap) at finite temperatures~\cite{garate2013,saha2014}, while Antonius and Louie further provided a symmetry argument making the picture more nuanced~\cite{antonius2016}.

\section{Temperature-tunable topological states in 1T$^\prime$-WS$_2$}
\label{sec: v}
The strong electron-phonon coupling in 1T$^\prime$-WS$_2$ which promotes topology offers a facile mechanism to control the topological order by temperature. To realise this, we consider applying positive biaxial strain to first drive the system to a normal insulator with an uninverted gap. Without considering electron-phonon coupling, we find that the inverted band gap first decreases to zero at strain of 2.2\% and then reopens as the strain increases. By explicitly calculating the $Z_2$ topological invariant, we confirm that this gap-closing indeed induces a topological phase transition associated with the destruction of the edge states.

\begin{figure}
\centering
\includegraphics[width=1.0\linewidth]{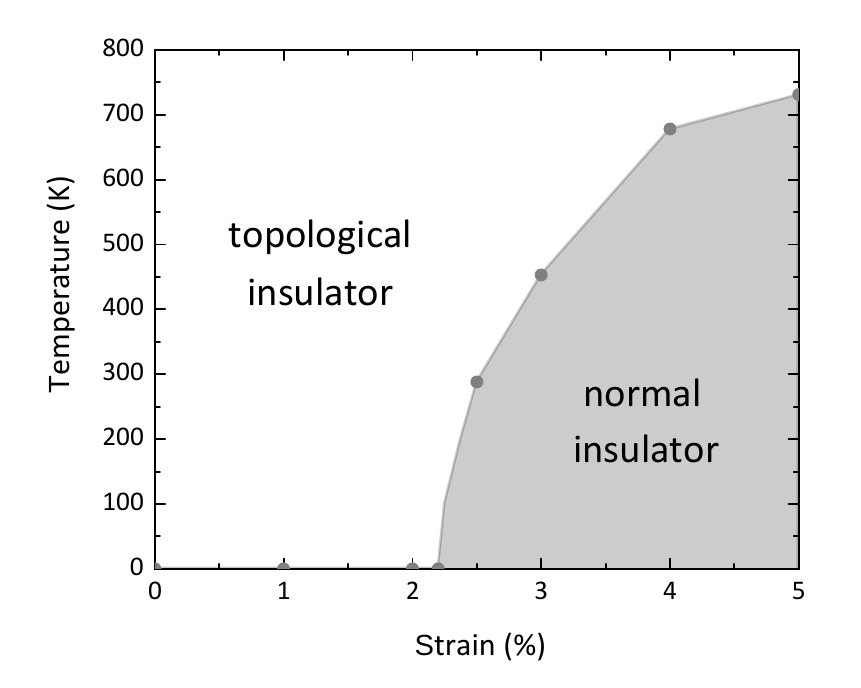}  
\caption{Temperature-strain phase diagram of WS$_2$.}
\label{fig:band_vs_strain}
\end{figure}

Figure\,\,\ref{fig:band_vs_strain} shows the temperature-strain phase diagram of WS$_2$. In the range of strain up to 5\%, we consistently observe that electron-phonon coupling drives the electronic structure of 1T$^\prime$-WS$_2$ toward its topologically non-trivial region. This agrees with the previous theoretical model introduced in Section ~\ref{sec: iv}. In particular, we find that at strain of approximately 2.5\%, elevating the temperature to $300$\,K drives a phase transition taking the system from the topologically trivial state imposed by strain to the topologically non-trivial phase. As the first example of temperature promoted topological insulating phases, this temperature-strain tunable state holds promise for tailoring device functionalities.

\section{Conclusion}
\label{sec: vi}
In summary, through first principles calculations, we have elucidated the role of three critical temperature-related factors: electron-phonon coupling, Fröhlich coupling, and thermal expansion, providing a comprehensive examination of the intricate temperature effects in 1T$^\prime$-MX$_2$ monolayers.

Our findings demonstrate that within 1T$^\prime$-MX$_2$ monolayers electron-phonon coupling generally promotes the topology of the electronic structures. However, the counteracting influence of thermal expansion should not be overlooked, as it generally diminishes the topological attributes and holds the potential to reverse the temperature dependence of the band gap in some cases. This finding also sheds light on using a substrate that can suppress thermal expansion to achieve better thermal robustness of 2D topological insulators.

Furthermore, our investigation into Fröhlich coupling in 2D materials has revealed its relatively modest temperature dependence within 1T$^\prime$-MX$_2$ monolayers due to the small ionic contribution to the dielectric constant. However, the formulation we have found is universal for all 2D materials, which can be useful for understanding the strong band renormalization in the 2D materials that exhibit large ionic contribution to the dielectric constant.

In the context of material science, one of the outcomes of our study is the identification of MoSe$_2$ as a promising candidate for room temperature applications. It exhibits remarkable resilience against thermal expansion, making it a robust choice for electronic devices operating at higher temperatures. Additionally, WS$_2$ displays tunable topological behavior under the combined influence of strain and temperature, opening up possibilities for tailored device functionalities. Both materials stand out as novel examples of temperature promoted topological insulators.

Overall, our work advances the fundamental understanding of temperature effects in 1T$^\prime$-MX$_2$ monolayers, paving the way for the applications of 2D topological insulators.

\begin{acknowledgments}
S.C. acknowledges financial support from the Cambridge Trust and from the Winton Programme for the Physics of Sustainability. I.J.P. acknowledges financial support from the Worshipful Company of Armourers and Brasiers, Mr Peter Mason, and the Henry Royce Institute. I.J.P. also acknowledges support from Dr Andrew Spencer through the Gonville and Caius College Senior Tutor's Internship grant. B.M. acknowledges support from a UKRI Future Leaders Fellowship (MR/V023926/1), from the Gianna Angelopoulos Programme for Science, Technology, and Innovation, and from the Winton Programme for the Physics of Sustainability. The calculations have been performed using resources provided by the Cambridge Tier-2 system (operated by the University of Cambridge Research Computing Service and funded by EPSRC [EP/P020259/1]), as well as by the UK Materials and Molecular Modelling Hub (partially funded by EPSRC [EP/P020194]), Thomas, and by the UK National Supercomputing Service, ARCHER. Access to Thomas and ARCHER was obtained via the UKCP consortium and funded by EPSRC [EP/P022561/1]).  
\end{acknowledgments}

\appendix

\section{Real valued phonon displacement operator}
\label{appendixa}

Following the procedure that has been described in many textbooks of lattice dynamics~\cite{born1955,dove1993}, we define the normal coordinates $u_{\nu \boldsymbol{q}}$ from the atomic displacements ${l}_{p \alpha j}$ for the phonon system as follows:
\begin{equation}
\label{Eq.{munuq}}
u_{\nu \boldsymbol{q}}=\frac{1}{\sqrt{N_{\mathrm{p}}}} \sum_{p \alpha j} \sqrt{m_{\alpha}} l_{p \alpha j} \mathrm{e}^{-\mathrm{i} \boldsymbol{q} \cdot \boldsymbol{\mathcal{R}}_{p \alpha}} \xi^\ast_{\nu \boldsymbol{q}  \alpha j},
\end{equation}
or inversely, 
\begin{equation}
\label{Eq.{lpaj}}
l_{p \alpha j}=\frac{1}{\sqrt{N_{\mathrm{p}}}} \sum_{\nu \boldsymbol{q}} \frac{1}{\sqrt{m_{\alpha}}}u_{\nu \boldsymbol{q}}\mathrm{e}^{\mathrm{i} \boldsymbol{q} \cdot \boldsymbol{\mathcal{R}}_{p \alpha}} \xi_{\nu \boldsymbol{q}  \alpha j},
\end{equation}
where $p$ and $\alpha$ run over all primitive cells of the crystal and the ions in the cell respectively, $j$ denotes the Cartesian components, $m_\alpha$ is the mass of the $\alpha$-th ion, $N_\mathrm{p}$ is the number of primitive cells in the crystal, and $\xi_{\nu \boldsymbol{q} \alpha j}$ is the eigenvector of the dynamical matrix of the system.

It is worth noting that $u_{\nu \boldsymbol{q}}$ is generally complex (because $\xi_{\nu \boldsymbol{q}  \alpha j}$ is complex), but one can construct a set of real-valued phonon displacement from $ u_{\nu \boldsymbol{q}}$ by partitioning the phonon BZ into three parts:
\begin{equation}
\mathrm{BZ} = \mathrm{BZ}_0 \cup \mathrm{BZ}_{+1} \cup \mathrm{BZ}_{-1}.
\end{equation}
Here $\mathrm{BZ}_0 = \{\boldsymbol{q} \notin \mathrm{BZ}_{\pm1} \,|\, \boldsymbol{q} = \boldsymbol{-q} \; \mathrm{mod} \; \mathcal{G}\}$ represents a set of discrete $\boldsymbol{q}$ points invariant under inversion modulo a reciprocal lattice vector (\ie the centre of the Brillouin zone, the centres of its faces, and the corners), and  $\mathrm{BZ}_{\pm1} = \{\boldsymbol{q} \notin \mathrm{BZ}_0 \,|\, -\boldsymbol{q} \in \mathrm{BZ_{\mp1}} \}$ are mutually inversion symmetric images, each including all the $\boldsymbol{q}$-points that are not inversion partners. On this partitioning, we define the real phonon displacement 
\begin{equation}
\label{Eq.{munuq2}}
\mu_{\nu \boldsymbol{q}}= \begin{dcases}u_{\nu \boldsymbol{q}} & \text { for } \boldsymbol{q} \in \mathrm{BZ}_0 \\ \; \frac{1}{\sqrt{2}}\left({u}_{\nu \boldsymbol{q}}+{u}_{\nu \boldsymbol{-q}}\right) & \text { for } \boldsymbol{q} \in \mathrm{BZ}_{+1}\\ \; \frac{\mathrm{i}}{\sqrt{2}}\left({u}_{\nu \boldsymbol{q}}-{u}_{\nu \boldsymbol{-q}}\right) & \text { for } \boldsymbol{q} \in \mathrm{BZ}_{-1} \end{dcases}, 
\end{equation}
or inversely, 
\begin{equation}
u_{\nu \boldsymbol{q}}= \begin{dcases}\mu_{\nu \boldsymbol{q}} & \text { for } \boldsymbol{q} \in \mathrm{BZ}_0 \\ \; \frac{1}{\sqrt{2}}\left({\mu}_{\nu \boldsymbol{q}}+\mathrm{i}{\mu}_{\nu \boldsymbol{-q}}\right) & \text { for } \boldsymbol{q} \in \mathrm{BZ}_{+1}\\ \; \frac{1}{\sqrt{2}}\left({\mu}_{\nu \boldsymbol{-q}}-\mathrm{i}{\mu}_{\nu \boldsymbol{q}}\right) & \text { for } \boldsymbol{q} \in \mathrm{BZ}_{-1} \end{dcases}. 
\end{equation}
From Eq.\,\eqref{Eq.{munuq2}}, one can naturally define a real-value phonon differential operator $\frac{\partial}{\partial \mu_{\nu \boldsymbol{q}}}$, which has been used in the main text to reformulate AHC theory.

We note that one widely used formulation of $E_{n\boldsymbol{k}}(T)$ is given as follows~\citep{giustino2010}\footnote{The sum excludes $\boldsymbol{q}=\boldsymbol{0}$ for the first term in $\sum_{\nu \boldsymbol{q}}[\ldots]$ when $m=n$.}:
\begin{equation} 
\label{Eq.{E_nk(T)-g}}
\begin{aligned}
&E_{n\boldsymbol{k}}(T)=E_{n\boldsymbol{k}}^{(0)} + \\ &\frac{1}{N_{\mathrm{p}}}\sum_{\nu  \boldsymbol{q}} \left[\sum_{m} \frac{\left|g_{m n \nu}(\boldsymbol{k}, \boldsymbol{q})\right|^{2}}{E_{n\boldsymbol{k}}^{(0)} -E_{m \boldsymbol{k}+\boldsymbol{q}}^{(0)}}+g_{n n \nu \nu}^{\mathrm{DW}}(\boldsymbol{k}, \boldsymbol{q},-\boldsymbol{q})\right] \\ &\times \left[1+2n_{\mathrm{B}}\left(\omega_{\nu \boldsymbol{q}}, T\right)\right],
\end{aligned}
\end{equation}
where 
\begin{equation}
\label{Eq.{g_mnv}}
g_{m n \nu}(\boldsymbol{k}, \boldsymbol{q})=\left\langle \varphi_{m \boldsymbol{k}+\boldsymbol{q}}\left|\partial_{\nu \boldsymbol{q}} \mathcal{V}_{\mathrm{el}} \right| \varphi_{n \boldsymbol{k}}\right\rangle
\end{equation}
is referred to as the standard electron-phonon coupling matrix element, $\mathcal{V}_{\mathrm{el}}$ is the potential experienced by the electrons in crystals, and $g_{n n \nu \nu}^{\mathrm{DW}}(\boldsymbol{k}, \boldsymbol{q},-\boldsymbol{q})$ is a particular case of the Debye-Waller electron-phonon matrix 
 \begin{equation}
  g_{m n \nu \nu^{\prime}}^{\mathrm{DW}}\left(\boldsymbol{k}, \boldsymbol{q}, \boldsymbol{q}^{\prime}\right) =  \frac{1}{2}\left\langle \varphi_{m \boldsymbol{k}+\boldsymbol{q}+\boldsymbol{q}^{\prime}}\left|\partial_{\nu \boldsymbol{q}} \partial_{ \nu^{\prime} \boldsymbol{q}^{\prime}}  \mathcal{H}_{\mathrm{el}} \right| \varphi_{n \boldsymbol{k}}\right\rangle.
 \end{equation}
Here the (complex) phonon differential operator $\partial_{\nu \boldsymbol{q}}$ is formulated as~\citep{giustino2007, giustino2017}
\begin{equation}
\label{Eq.{partial_nuq}}
\partial_{\nu \boldsymbol{q}} \equiv \frac{1}{\sqrt{2 \omega_{\nu \boldsymbol{q}}}} \sum_{p \alpha j} \frac{1}{\sqrt{m_{\alpha}}} \mathrm{e}^{\mathrm{i} \boldsymbol{q} \cdot \boldsymbol{\mathcal{R}}_{p \alpha}} \xi_{\nu \boldsymbol{q}  \alpha j} \frac{\partial}{\partial l_{p \alpha j}}.
\end{equation}
We show that Eq.\,\eqref{Eq.{Enk(T)-quadratic}} and Eq.\,\eqref{Eq.{E_nk(T)-g}} are equivalent. First, the second derivative term in Eq.\,\eqref{Eq.{Enk(T)-quadratic}} can be split into two terms by invoking the Hellmann-Feynman theorem~\citep{deb1972}, \ie
\begin{equation}
\begin{aligned}
\label{Eq.{HF-split}}
\frac{\partial^{2} E_{n \boldsymbol{k}}}{\partial \mu_{\nu \boldsymbol{q}}^{2}} = &\left\langle\varphi_{n \boldsymbol{k}}\left|\frac{\partial^{2} \mathcal{H}_{\mathrm{el}}}{\partial \mu_{\nu \boldsymbol{q}}^{2}}\right| \varphi_{n \boldsymbol{k}}\right\rangle + \\ &\left\langle\varphi_{n \boldsymbol{k}}\left|\frac{\partial\mathcal{H}_{\mathrm{el}}}{\partial \mu_{\nu \boldsymbol{q}}}\right| \frac{\partial \varphi_{n \boldsymbol{k}}}{\partial \mu_{\nu \boldsymbol{q}}}\right\rangle+\left\langle \frac{\partial \varphi_{n \boldsymbol{k}}}{\partial \mu_{\nu \boldsymbol{q}}}\left|\frac{\partial\mathcal{H}_{\mathrm{el}}}{\partial \mu_{\nu \boldsymbol{q}}}\right|\varphi_{n \boldsymbol{k}}\right\rangle,
\end{aligned}
\end{equation}
where the derivative of the state can be resolved by an unperturbed complete basis set $\{|\varphi_{n \boldsymbol{k}}\rangle\}$ according to perturbation theory~\citep{baroni2001} as follows:
\begin{equation}
\label{Eq.{dphi/dmu}}
\left| \frac{\partial \varphi_{n \boldsymbol{k}}}{\partial \mu_{\nu \boldsymbol{q}}}\right\rangle =\sum_{\tiny{\begin{array}{c}(m, \boldsymbol{k^\prime})\\ \neq  (n, \boldsymbol{k}) \end{array}}}\left|\varphi_{m \boldsymbol{k}^\prime}\right\rangle \frac{\left\langle\varphi_{m \boldsymbol{k}^\prime}\left|\frac{\partial\mathcal{H}_{\mathrm{el}}}{\partial \mu_{\nu \boldsymbol{q}}}\right| \varphi_{n \boldsymbol{k}}\right\rangle}{E_{n \boldsymbol{k}}^{(0)} -E_{m \boldsymbol{k}^\prime}^{(0)} }.
\end{equation}
Second, using the chain rule, one can find that the complex phonon differential operator is related to our real-displacement phonon differential operator by the relation
\begin{equation}
\frac{\partial}{\partial \mu_{\nu \boldsymbol{q}}}= \begin{dcases} \sqrt{\frac{2\omega_{\nu \boldsymbol{q}}}{N_\mathrm{p}}}\partial_{\nu \boldsymbol{q}} & \text { for } \boldsymbol{q} \in \mathrm{BZ}_0 \\ \; 2\sqrt{\frac{\omega_{\nu \boldsymbol{q}}}{N_\mathrm{p}}}\partial_{\nu \boldsymbol{q}} & \text { for } \boldsymbol{q} \in \mathrm{BZ}_{+1}\\ \; 0 & \text { for } \boldsymbol{q} \in \mathrm{BZ}_{-1} \end{dcases}, 
\end{equation}
and thus 
\begin{equation}
\label{Eq.{sum_dmu_vq}}
\lim_{N_\mathrm{p} \to \infty} \sum_{\nu\boldsymbol{q}} \frac{\partial}{\partial \mu_{\nu \boldsymbol{q}}} = \lim_{N_\mathrm{p} \to \infty} \sum_{\nu\boldsymbol{q}} \sqrt{\frac{\omega_{\nu \boldsymbol{q}}}{N_\mathrm{p}}}\partial_{\nu \boldsymbol{q}}.
\end{equation}
Substituting Eqs.\eqref{Eq.{HF-split}} - \eqref{Eq.{sum_dmu_vq}} into Eq.\,\eqref{Eq.{Enk(T)-quadratic}} yields exactly Eq.\,\eqref{Eq.{E_nk(T)-g}}, so the equivalence is proved.  We also note that Refs.\,\citep{ponce2014, ponce2015} derived another formulation of $E_{n\boldsymbol{k}}(T)$ using a generalized Janak’s theorem. Similarly, one can also show that their result 
\begin{equation}
\label{Eq.{E_nk(T)-nb}}
E_{n\boldsymbol{k}}(T)=E_{n\boldsymbol{k}}^{(0)} +\frac{1}{2N_\mathrm{p}} \sum_{ \nu \boldsymbol{q}}  \frac{\partial E_{n \boldsymbol{k}}}{\partial n_\mathrm{B}}\left[1+2n_{\mathrm{B}}\left(\omega_{\nu \boldsymbol{q}}, T\right)\right]
\end{equation}
with
\begin{equation}
\label{Eq.{dEnk/dnb}}
\begin{aligned}
\frac{\partial E_{n \boldsymbol{k}}}{\partial n_\mathrm{B}}= &\frac{1}{2 \omega_{\nu \boldsymbol{q}}}  \sum_{p \alpha j} \sum_{p^{\prime} \alpha^{\prime} j^{\prime}} \frac{\partial^{2} E_{n \boldsymbol{k}}}{\partial l_{p  \alpha j} \partial l_{p^{\prime} \alpha^{\prime} j}}  \mathrm{e}^{-\mathrm{i} \boldsymbol{q} \cdot\left(\boldsymbol{\mathcal{R}}_{p\alpha}-\boldsymbol{\mathcal{R}}_{p^{\prime}\alpha^{\prime}}\right)}\\ & \times \frac{\xi_{\nu \boldsymbol{q}  \alpha j}}{\sqrt{m_\alpha}} \frac{\xi^\ast_{\nu \boldsymbol{q}  \alpha^\prime j^\prime}}{\sqrt{m_{\alpha^\prime}}}
\end{aligned}
\end{equation}
is equivalent to ours. 

It is worth noting that the Debye-Waller term involving second-order electron-phonon matrix elements is very challenging to calculate in the density functional perturbation theory framework, therefore one has to invoke the rigid-ion approximation to rewrite it as the product of first-order electron-phonon matrix elements. However, it can be easily included in the finite difference framework as used in this work.

\section{Fröhlich Coupling in 2D}
\label{appendixb}
For polar insulators, it has been known that the presence of Fröhlich coupling can play a role in the additional renormalization of the band structure. Here we provide a derivation for this. We start by considering a hole at the $\Gamma$ point of the conduction band $E_{m\boldsymbol{q}}^{(0)}$ interacting with a single dispersionless polar longitudinal-optical phonon of frequency $\omega_{\mathrm{LO}}$. At a finite temperature $T$, the self-energy for the hole polaron reads 
\begin{equation}
\begin{aligned}
\Sigma^{\mathrm{Fr}}(\omega)= & \int \frac{\mathrm{d} \boldsymbol{q}}{\Omega_{\mathrm{BZ}}} |g(\boldsymbol{q})|^2 \\
& \times\left[\frac{1-n_{\mathrm{F}}(E_{\mathrm{c}\boldsymbol{q}}^{(0)}, T)+n_{\mathrm{B}}\left(\omega_{\mathrm{LO}}, T\right)}{\omega-E_{\mathrm{c}\boldsymbol{q}}^{(0)} -\omega_{\mathrm{LO}}+\mathrm{i} \eta}\right. \\
& \left.+\frac{n_{\mathrm{F}}(E_{\mathrm{c}\boldsymbol{q}}^{(0)}, T)+n_{\mathrm{B}}\left(\omega_{\mathrm{LO}}, T\right)}{\omega-E_{\mathrm{c}\boldsymbol{q}}^{(0)} +\omega_{\mathrm{LO}}+\mathrm{i} \eta}\right] .
\end{aligned}
\end{equation}
where $\Omega_\mathrm{BZ}$ is the volume of the Brillouin zone, $\eta$ is a positive infinitesimal,  $n_{\mathrm{F}}(E_{\mathrm{c}\boldsymbol{q}}^{(0)}, T) =[\exp(\frac{E_{\mathrm{c}\boldsymbol{q}}^{(0)}}{ k_{\mathrm{B}} T})+1]^{-1}$ is the Fermi–Dirac factor. The band renormalization arising from $\Sigma^{\mathrm{Fr}}(\omega)$ can be obtained from the standard prescription of many-body perturbation theory:   
\begin{equation}
\label{Eq.{polaron_re}}
\begin{aligned}
E_{\mathrm{c}\Gamma}(T)&=E_{\mathrm{c}\Gamma}^{(0)}+ \int \frac{\mathrm{d} \boldsymbol{q}}{\Omega_{\mathrm{BZ}}} \; |g(\boldsymbol{q})|^2 \\ & \times  \left[\frac{1+n_{\mathrm{B}}\left(\omega_{\mathrm{LO}}, T\right)}{-|\boldsymbol{q}|^2-q^2_{\mathrm{LO}}} +  \frac{n_{\mathrm{B}}\left(\omega_{\mathrm{LO}}, T\right)}{-|\boldsymbol{q}|^2+q^2_{\mathrm{LO}}}\right] ,
\end{aligned}
\end{equation}
where we have made the following approximations: (i)  the conduction band near the $\Gamma$ point is isotropic and parabolic, \ie $E_{\mathrm{c}\boldsymbol{q}}^{(0)} = \frac{\boldsymbol{q}^2}{2m^*}$, where $m^*$ is the effective mass; (2)  $k_{\mathrm{B}} T \ll E_{\mathrm{c}\boldsymbol{q}}^{(0)}$ thereby
$n_{\mathrm{F}}(E_{\mathrm{c}\boldsymbol{q}}^{(0)}, T) = 0$ for all conduction states near the $\Gamma$ point; (iii) $\Sigma^{\mathrm{Fr}}(\omega)$ is purely real-valued, \ie  $\mathrm{i}\eta = 0$. In addition, we have defined the effective longitudinal optical wavevector $q^2_{\mathrm{LO}} = 2m^*\omega_{\mathrm{LO}}$.

We first examine Eq.\,\eqref{Eq.{polaron_re}} in the 3D case, in which the Fröhlich electron-phonon coupling matrix has a long known expression given by~\cite{frohlich1950, frohlich1954, verdi2015}
\begin{equation}
\label{Eq.{frohlich_g}}
|g(\boldsymbol{q})|^2 = \frac{2\pi}{V_\mathrm{cell} }\omega_\mathrm{LO}\left(\frac{1}{\epsilon_{\infty}}-\frac{1}{\epsilon_0}\right) \frac{1}{|\boldsymbol{q}|^2},
\end{equation}
where $V_\mathrm{cell}$ is the volume of the primitive cell, $\epsilon_{\infty}$ and $\epsilon_{0}$ are the high-frequency and static relative permittivity. By substituting Eq.\,\eqref{Eq.{frohlich_g}} into Eq.\,\eqref{Eq.{polaron_re}} and approximating the Brillouin zone integration as $\int \mathrm{d} \boldsymbol{q} = \int_0^{q_{\mathrm{BZ}}} \mathrm{d} q \; 4\pi q^2 =\Omega_{\mathrm{BZ}}$, we arrive at
\begin{equation}
\begin{aligned}
E_{\mathrm{c}\Gamma}(T)&=E_{\mathrm{c}\Gamma}^{(0)}+ \int_0^{q_{\mathrm{BZ}}} {\mathrm{d} q} \; \frac{2}{\pi}m^*\omega_\mathrm{LO}\left(\frac{1}{\epsilon_{\infty}}-\frac{1}{\epsilon_0}\right) \\ & \times  \left[\frac{1+n_{\mathrm{B}}\left(\omega_{\mathrm{LO}}, T\right)}{-q^2-q^2_{\mathrm{LO}}} +  \frac{n_{\mathrm{B}}\left(\omega_{\mathrm{LO}}, T\right)}{-q^2+q^2_{\mathrm{LO}}}\right] \\
& = E_{\mathrm{c}\Gamma}^{(0)}+\frac{2}{\pi}m^*\omega_\mathrm{LO}\left(\frac{1}{\epsilon_{\infty}}-\frac{1}{\epsilon_0}\right) \\ & \times \left[-\frac{1+n_{\mathrm{B}}\left(\omega_{\mathrm{LO}}, T\right)}{q_{\mathrm{LO}}}\tan^{-1}\left(\frac{q_\mathrm{BZ}}{q_{\mathrm{LO}}}\right) \right. \\
& \left. +\frac{n_{\mathrm{B}}\left(\omega_{\mathrm{LO}}, T\right)}{q_{\mathrm{LO}}}\tanh^{-1}\left(\frac{q_\mathrm{LO}}{q_{\mathrm{BZ}}}\right)  \right].
\end{aligned}
\end{equation}
A frequently employed treatment to further simplify the above expression is to set $q_\mathrm{BZ} \to \infty$, typically leading to an error in the value of the integral of the order of 10\%~\cite{callaway1976}. Following this treatment and considering $T=0$, we obtain the Fröhlich zero-point band renormalization $\Delta E^\mathrm{Fr}_{\mathrm{c}\Gamma}(0) = -\frac{m^*\omega_{\mathrm{LO}}}{q_{\mathrm{LO}}}\left(\frac{1}{\epsilon_{\infty}}-\frac{1}{\epsilon_0}\right)  = - \alpha \omega_{\mathrm{LO}}$, where $\alpha = \frac{m^*}{q_{\mathrm{LO}}}\left(\frac{1}{\epsilon_{\infty}}-\frac{1}{\epsilon_0}\right)$ is the dimensionless polaron constant. This outcome aligns precisely with the previously known conclusion about polarons~\cite{callaway1976}. 

Now, we move to the 2D case. It is worth noting that unlike the 3D case where the volume element (\ie a spherical shell) $4 \pi q^2 \mathrm{d}q$ can alleviate the singularity of the integral at $q=0$, in the 2D case the area element (\ie an annulus) $2 \pi q \mathrm{d}q$ does not possess the same capability.  If one insists on using the Fröhlich electron-phonon coupling matrix given in Eq.~\eqref{Eq.{frohlich_g}} which is proportional to $|\boldsymbol{q}|^{-2}$, the 2D integral involved in Eq.\,\eqref{Eq.{polaron_re}} will invariably diverge. The primary cause for this divergence can be attributed to the oversimplification of a 2D material, treated merely as a mere sheet lacking thickness, in the modeling of polarons. This has been clearly described in Ref.~\cite{sio2023}. 

To avoid the divergence mentioned above, we employ the recently proposed 2D Fröhlich electron-phonon coupling matrix ~\cite{sio2022}
\begin{equation}
\label{Eq.{frohlich_g_2d}}
|g(\boldsymbol{q})|^2 = \left[\frac{\pi d}{2A_\mathrm{cell}} \omega_{\mathrm{LO}}\left(\epsilon_0-\epsilon_{\infty}\right)\right] \frac{q_0^2}{(|\boldsymbol{q}|+q_0)^2},
\end{equation}
to derive the Fröhlich band renormalization for 2D materials, where $A_\mathrm{cell}$ is the area of the primitive cell, $d$ is the effective thickness of the 2D material, and $q_0$ is defined as
\begin{equation}
\label{Eq.{q0}}
q_0=\frac{4 \epsilon_{\infty}}{(2 \epsilon_{\infty}^2-1)d}.
\end{equation}
Again, by substituting Eq.\,\eqref{Eq.{frohlich_g_2d}} into Eq.\,\eqref{Eq.{polaron_re}} and approximating the Brillouin zone integration as $\int \mathrm{d} \boldsymbol{q} = \int_0^\infty \mathrm{d} q \; 2\pi q$, we arrive at 

\begin{widetext}
\begin{equation}
\label{Eq.{frohlich_re_2d_ana}}
\begin{aligned}
E_{\mathrm{c}\Gamma}(T)&=E_{\mathrm{c}\Gamma}^{(0)}+ \int_0^{\infty} {\mathrm{d} q} \; \frac{1}{2}m^*q_0^2{d} \omega_{\mathrm{LO}}\left(\epsilon_0-\epsilon_{\infty}\right) \frac{q}{(q_0+q)^2} \left[\frac{1+n_{\mathrm{B}}\left(\omega_{\mathrm{LO}}, T\right)}{-q^2-q^2_{\mathrm{LO}}} +  \frac{n_{\mathrm{B}}\left(\omega_{\mathrm{LO}}, T\right)}{-q^2+q^2_{\mathrm{LO}}}\right] \\
& = E_{\mathrm{c}\Gamma}^{(0)}+\frac{1}{2}m^*q_0^2{d} \omega_{\mathrm{LO}}\left(\epsilon_0-\epsilon_{\infty}\right) \left\{-\frac{q_{\mathrm{LO}}^2-q_0^2 \left[2 n_{\mathrm{B}}\left(\omega_{\mathrm{LO}}, T\right) + 1\right]}{q_0^4-q_{\mathrm{LO}}^4} - \frac{\pi q_0 q_{\mathrm{LO}} \left[n_{\mathrm{B}}\left(\omega_{\mathrm{LO}}, T\right)+1\right]}{\left(q_0^2+q_{\mathrm{LO}}^2\right)^2} \right. \\
& \left. -\frac{\left[2 q_0^2 \left(q_0^4+3 q_{\mathrm{LO}}^4\right) n_{\mathrm{B}}\left(\omega_{\mathrm{LO}}, T\right)+\left(q_0^2-q_{\mathrm{LO}}^2\right)^3\right] \left[\ln\left(q_0\right)-\ln\left(q_{\mathrm{LO}}\right)\right]}{\left(q_0^4-q_{\mathrm{LO}}^4\right)^2}\right\}.
\end{aligned}
\end{equation}
\end{widetext}
In a similar manner, one can find the expression for the correction to the valence band $E_{\mathrm{v}\Gamma}(T)$. 

Invoking the adiabatic approximation by replacing the term in the bracket of Eq.\,\eqref{Eq.{frohlich_re_2d}} by $[2n_{\mathrm{B}}(\omega_\mathrm{LO},T)+1]/(-q^2 + \mathrm{i}2m^*\eta)$ and then taking the real part~\cite{nery2016}, we obtain:

\begin{widetext}
\begin{equation}
\label{frohlich_re_2d_ad}
\begin{aligned}
E_{\mathrm{c}\Gamma}(T)&=E_{\mathrm{c}\Gamma}^{(0)}+ \mathfrak{Re}\left\{ \int_0^{\infty} {\mathrm{d} q} \; \frac{1}{2}m^*q_0^2{d} \omega_{\mathrm{LO}}\left(\epsilon_0-\epsilon_{\infty}\right) \frac{q}{(q_0+q)^2} \frac{2n_{\mathrm{B}}(\omega_\mathrm{LO},T)+1}{-q^2 + \mathrm{i}2m^*\eta}\right\} \\
&=E_{\mathrm{c}\Gamma}^{(0)}+ \frac{1}{8}m^*q_0^2{d} \omega_{\mathrm{LO}}\left(\epsilon_0-\epsilon_{\infty}\right)\left[2n_{\mathrm{B}}(\omega_\mathrm{LO},T)+1\right] \\ &\times 
 \mathfrak{Re}\left\{-\frac{2(\pi-4\mathrm{i})m^*\eta-(4-4\mathrm{i})\pi q_0\sqrt{m^*\eta} }{(2m^*\eta + \mathrm{i}q_0^2)^2} - \frac{(4-\mathrm{i}\pi)q_0^2+2(q_0^2 + \mathrm{i} 2m^*\eta)\left[\ln(2m^*\eta/q_0^2)\right]}{(2m^*\eta + \mathrm{i}q_0^2)^2}  \right\}
\end{aligned}
\end{equation}
\end{widetext}

It is worth noting that the small imaginary component $i\eta$ introduced here is \textit{ad hoc} rather than \textit{ab initio}. The physical meaning of $\eta$ can be inferred as a finite lifetime for the unoccupied electronic states due to thermal effects. In principle, in more accurate approaches the $\eta$ should be replaced by the finite physical linewidth of electrons. As already pointed out by Ref.~\cite{ponce2015}, decreasing $\eta$ does not lead to convergence for polar materials. One has to treat this with caution in band renormalization calculations.


\bibliography{references}
\end{document}